\documentclass[graybox,b5paper]{svmult}

\usepackage{mathptmx}       
\usepackage{helvet}         
\usepackage{courier}        
\usepackage{type1cm}        
\usepackage{makeidx}         
\usepackage{graphicx}        
\usepackage{multicol}        
\usepackage[bottom]{footmisc}

\makeindex             

\usepackage{url}

\usepackage{ifthen}
\usepackage{xifthen}
\usepackage{ifthenx}

\usepackage{color}
\usepackage{times}
\usepackage{txfonts}
\usepackage{bm}

\newcommand{\ShortVersion}[1]{#1}
\newcommand{\LongVersion}[1]{}

\newcommand{\subsubsubsection}[1]{\vspace*{1em}\newline\noindent\textit{#1}\vspace*{1em}\noindent}
\newcommand{\script}[1]{{\mbox{\scriptsize #1}}}

\newcommand{\CITE}[1]{\cite{#1}}
\newcommand{\VEC}[1]{\text{\boldmath{$#1$}}}

\newcommand{\Eqno}[1]{{#1}}
\newcommand{\Ref}[1]{\Eqno{\ref{#1}}}
\newcommand{\Eq}[1]{Eq. \Eqno{#1}}
\newcommand{\Eqs}[1]{Eqs. {#1}}

\newcommand{\Fig}[1]{Fig. {#1}}

\newcommand{\Table}[1]{Table {#1}}

\newcommand{\FigureInText}[1]{}

\newcommand{\FigureInLegends}[1]{}

\newcommand{\TableInText}[1]{}
\newcommand{\TableLegends}[1]{}
\newcommand{\TableInLegends}[1]{#1}

\newcommand{\EQ}{eq}
\newcommand{\TBL}{tbl}
\newcommand{\FIG}{fig}
\renewcommand{\CITE}[1]{\cite{#1}}

\renewcommand{\subsubsubsection}[1]{\vspace*{1em}\noindent{#1}\vspace*{1em}\noindent}
\renewcommand{\subsubsubsection}[1]{\paragraph{#1}}

\newcommand{\text}[1]{\textrm{#1}}
\renewcommand{\VEC}[1]{\text{\boldmath{$#1$}}}

\ifdefined\text
\else
\newcommand{\text}[1]{\textrm{#1}}
\fi

\ifdefined\EQ
\renewcommand{\EQ}{eq}
\else
\newcommand{\EQ}{eq}
\fi
\ifdefined\TBL
\renewcommand{\TBL}{tbl}
\else
\newcommand{\TBL}{tbl}
\fi
\ifdefined\FIG
\renewcommand{\FIG}{fig}
\else
\newcommand{\FIG}{fig}
\fi

\renewcommand{\FigureInText}[1]{}

\renewcommand{\FigureInLegends}[1]{#1}
\newcommand{\NoFigureInLegends}[1]{#1}
\FigureInLegends{
\renewcommand{\NoFigureInLegends}[1]{}
}

\newcommand{\NoFigureInText}[1]{#1}
\FigureInText{
\renewcommand{\NoFigureInText}[1]{}
}

\renewcommand{\TableInText}[1]{}

\renewcommand{\TableInLegends}[1]{#1}
\newcommand{\NoTableInLegends}[1]{#1}
\TableInLegends{
\renewcommand{\NoTableInLegends}[1]{}
}

\renewcommand{\TableLegends}[1]{#1}
\newcommand{\NoTableInText}[1]{#1}
\TableInText{
\renewcommand{\NoTableInText}[1]{}
}

\newcommand{\SupFig}[1]{}

\newcommand{\SupTable}[1]{}

\newcommand{\SupplementaryBibliography}[1]{#1}

\newcommand{\TextMaterial}[1]{#1}
\TextMaterial{
\renewcommand{\SupplementaryBibliography}[1]{}
}

\ifdefined\SUPPLEMENT
\else
\newcommand{\SUPPLEMENT}[1]{}
\fi
\ifdefined\TEXT
\else
\newcommand{\TEXT}[1]{#1}
\fi

\FigureInText{
}
\TableInText{
}

\renewcommand{\ShortVersion}[1]{}
\renewcommand{\LongVersion}[1]{#1}

\begin{document}

\title*{Prediction of Structures and Interactions from Genome Information}
\titlerunning{Structure Prediction from Genome Information}
\author{Miyazawa, Sanzo}
\institute{Sanzo Miyazawa, \email{sanzo.miyazawa@gmail.com}}
%
%
\maketitle

\abstract{
Predicting three dimensional residue-residue contacts
from evolutionary information in protein sequences was attempted already
in the early 1990s.
However,
contact prediction accuracies 
of methods evaluated in CASP experiments before CASP11
remained quite low,
typically with 
$<20$\% true positives. 
Recently,  
contact prediction
has been significantly improved to the level that
an accurate three dimensional model
of a large 
protein
can be generated on the basis of predicted contacts.
This improvement was attained by 
disentangling direct from indirect correlations 
in amino acid covariations or cosubstitutions between sites
in protein evolution.
Here, we review statistical methods for extracting 
causative correlations
and various approaches to describe protein structure, complex, and flexibility 
based on predicted contacts.
} 
\vspace*{2em}
\noindent
Keywords: contact prediction; direct coupling; amino acid covariation; amino acid cosubstitution; 
partial correlation; maximum entropy model; inverse Potts model; Markov random field; 
Boltzmann machine; deep neural network

\section{Introduction}

\begin{figure}[hbt]
\centerline{
\includegraphics*[width=75mm,angle=0]{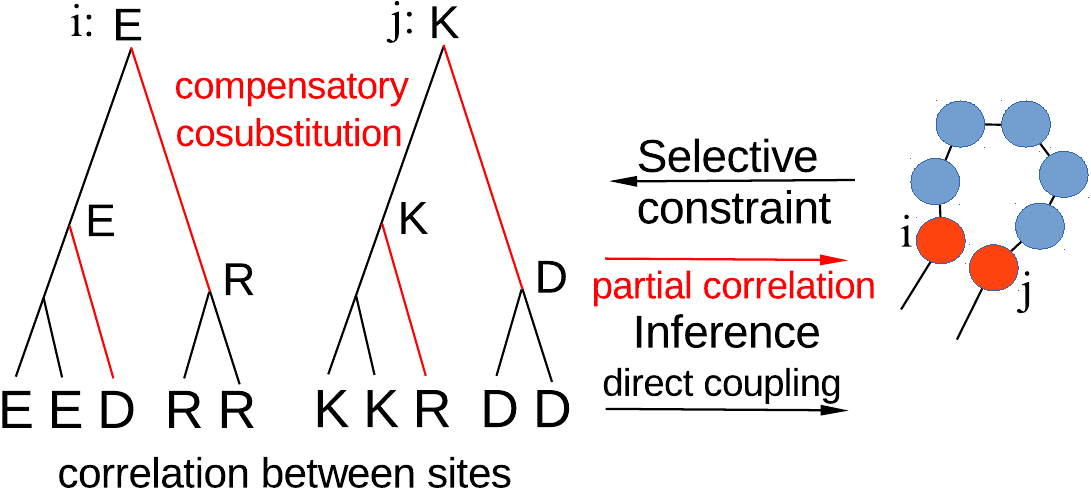}
}
\caption
{
\label{fig: schema}
\noindent
Amino acids at sites $i$ and $j$ in a MSA are shown with a phylogenetic tree. 
Causative correlations
between sites in protein evolution are extracted 
from the MSA or phylogenetic tree, 
and utilized to infer close residue pairs.
}
\end{figure}

The evolutionary history of protein sequences
is a valuable source of information in many fields of science
not only in evolutionary biology but even to understand protein structures.
Residue-residue interactions that fold a protein into a unique three-dimensional (3D) structure
and make it play a specific function 
impose structural and functional constraints 
in varying degrees 
on each amino acid.
Selective constraints on amino acids are recorded
in amino acid orders in homologous protein sequences
and also in the evolutionary trace of amino acid substitutions.
Negative effects caused by mutations at one site 
must be compensated by successive mutations at other sites
\CITE{YHT:64,FM:70,MA:04},
causing covariations/cosubstitutions/coevolution between sites\CITE{TD:00,FYB:04,DPJG:05,DG:07},
otherwise most negative mutants 
will be eliminated from a gene pool and never reach fixation in population.
Such structural and functional constraints arise from interactions between sites 
mostly in close spatial proximity. 
Thus, it has been suggested and also shown
that the types of amino acids  
\CITE{LGLS:99,LGJ:02,LGJ:12,RLMYR:05,SPSLAGL:08,BN:08,WWSHH:09,HRLR:09,BN:10,MPLBMSZOHW:11,MCSHPZS:11}
and amino acid substitutions 
\CITE{AVBMN:88,GSSV:94,SKS:94,PT:97,PTG:99,AWFTD:00,FOVC:01,FA:04,FYB:04,DPJG:05,MGDW:05,FT:06,DP:07,DG:07,DWG:08,PLFP:08,D:12,G-K:12}
are correlated between sites that are close in a protein 3D structure.
However, until CASP11,
contact prediction accuracy remained quite low,
typically with 
$\leq 20$\% true positives
for top-$L/5$ long-range contacts in free modeling targets\CITE{KJ:16}; $L$ denotes protein length.
Recently
contact prediction has been significantly improved to the level that an accurate
three dimensional model of a large protein ($\simeq 250$ residues) can be generated on the basis of
predicted contacts\CITE{CASP11:16}.
These improvements were attained primarily 
by disentangling direct from indirect correlations
in amino acid covariations or cosubstitutions
between sites in protein evolution, and secondarily by reducing
phylogenetic biases in a multiple sequence alignment (MSA) or removing them on the basis of a phylogenetic tree;
see \Fig{\ref{fig: schema}}.

Here, we review statistical methods for extracting 
causative correlations in amino acid covariations/cosubstitutions
between sites, and various approaches 
to describe protein structure, complex and flexibility
based on predicted contacts.
Mathematical formulation of each statistical method is concisely described in the unified manner 
\LongVersion{
in an appendix.
This manuscript was published in \CITE{M:18} with the short version of
the appendix.
} 
\ShortVersion{
in an appendix, the full version of which will be found in the article\CITE{M:17b} submitted to arXiv.
} 

\section{Statistical methods to extract causative interactions between sites}
\index{contact prediction}

The primary task to develop a robust method toward contact prediction
is to detect causative correlations, which reflect evolutionary constraints,
in amino acid covariations between sites in a multiple sequence alignment (MSA) 
or in amino acid cosubstitutions between sites in branches of a phylogenetic tree;
see \Table{\ref{tbl: basic_methods}}.
The former was called direct coupling analysis (DCA)\CITE{MPLBMSZOHW:11}.

\begin{table}[!ht]
\caption{
\label{tbl: basic_methods}
Statistical methods for disentangling direct from indirect correlations between sites.
} 

\vspace*{1em}
\footnotesize
\begin{tabular}{lll}
\hline
\multicolumn{2}{l}{Category} \\
 & Method name	&	Method/algorithm
\\
\hline	
\multicolumn{3}{l}{A) Direct coupling analysis of amino acid covariations between sites in a MSA}	\\
\hspace*{1em} 
& Boltzmann machine	& Markov chain Monte Carlo to calculate marginal probabilities and 
			\\
&	& gradient descent to estimate fields and couplings.
	\\
& CMI \CITE{LGJ:12}	& Boltzmann machine to estimate conditional mutual information 
	\\
& mpDCA \CITE{WWSHH:09}	
	& Message-passing algorithm to estimate marginal probabilities \\
&	& and gradient descent to estimate fields and couplings \\
& mfDCA \CITE{MPLBMSZOHW:11,MCSHPZS:11}	
	& Mean field approximation to estimate the partition function	\\
& PSICOV \CITE{JBCP:12}	
	& Graphical lasso (Gaussian approximation with an exponential prior) 	\\
&	& with a shrinkage method for a covariance matrix
	\\
& GaussDCA \CITE{BZFPZWP:14}
	& A multivariate Gaussian model with
	a normal-inverse-Wishart prior \\
& plmDCA \CITE{ELLWA:13,EHA:14}
	& Pseudo-likelihood maximization with Gaussian priors ($\ell_2$ regularizers)	\\
& GREMLIN \CITE{BKCLL:11,KOB:13}	
	& Pseudo-likelihood maximization with $\ell_1$ regularization terms\CITE{BKCLL:11} 
	\\
&	& or with Gaussian priors\CITE{KOB:13} which depend on site pair	\\
& ACE \CITE{CM:11,CM:12,BLCC:16}	
	& Adaptive cluster expansion of cross-entropy with Gaussian priors	\\
& Persistent VI \& Fadeout
	& Variational inference with sparsity-inducing prior, horseshoe \CITE{IM:16}
	\\
& \CITE{SMVG:15}
	& Boltzmann machine with $\ell_2$ regularization terms 	
	\\
& DI \CITE{TS:11}	
	& Partial correlation of normalized mutual informations between sites	\\ 
	\\
\multicolumn{3}{l}{B) Partial correlation analysis of amino acid cosubstitutions between sites in a phylogenetic tree}	\\
& pcSV\CITE{M:13}	
	& Partial correlation coefficients of coevolutionary substitutions 
	\\
&	& between sites within branches in a phylogenetic tree	\\
\hline
\end{tabular}

\vspace*{1em}
\noindent
\end{table}
\subsection{Direct coupling analysis for amino acid covariations between sites in a 
multiple sequence alignment}

The direct coupling analysis 
is based on the maximum entropy model for the distribution of protein sequences,
which satisfies the observed statistics in a MSA.

\subsubsection{Maximum entropy model for the distribution of protein sequences}
\index{contact prediction!maximum entropy model}
\index{maximum entropy model for contact prediction|see{contact prediction!maximum entropy model}}

Let us consider probability distributions $P(\VEC{\sigma})$ 
of amino acid sequences, $\VEC{\sigma} \equiv (\sigma_1, \ldots, \sigma_L)^T$
with $\sigma_i \in \{ \text{amino acids, deletion} \}$,
single-site and two-site marginal probabilities
of which are equal to a given frequency $P_i(a_k)$
of amino acid $a_k$ at each site $i$ and a given frequency $P_{ij}(a_k,a_l)$ of
amino acid pair $(a_k,a_l)$ for site pair $(i,j)$, respectively.
\begin{eqnarray}
	P(\sigma_i = a_k) &\equiv& 
	\sum_{\VEC{\sigma}} P(\VEC{\sigma}) \delta_{\sigma_i a_k} = P_i(a_k)
	\label{eq: constraints_single_site_marginals}
	\\
	P(\sigma_i = a_k, \sigma_j = a_l) &\equiv&
	\sum_{\VEC{\sigma}} P(\VEC{\sigma}) \delta_{\sigma_i a_k} \delta_{\sigma_j a_l} = P_{ij}(a_k, a_l)
	\label{eq: constraints_two_site_marginals}
\end{eqnarray}
where $a_k  \in \{ \text{amino acids, deletion} \}$,
$k = 1, \ldots, q$, 
$q \equiv | \{ \text{amino acids, deletion} \} | = 21$,
$i, j = 1, \ldots, L$,
and $\delta_{\sigma_i a_k}$ is the Kronecker delta. 
The distribution $P_{\script{ME}}$ with the maximum entropy is
\begin{eqnarray}
\lefteqn{
	P_{\script{ME}}(\VEC{\sigma}|h, J)
} 
	\\
 &=&
	\arg \max_{P(\VEC{\sigma})} [
		- \sum_{\VEC{\sigma}} P(\VEC{\sigma}) \log P(\VEC{\sigma})
		+ \lambda (\sum_{\VEC{\sigma}} P(\VEC{\sigma}) - 1)
		\nonumber \\
	& &
	+ \sum_i  \, [ \,  h_i(a_k) (\sum_{\VEC{\sigma}} P(\VEC{\sigma}) \delta_{\sigma_i a_k} - P_i(a_k)) \, ] \,
		\nonumber \\
	& &
	+ \sum_i \sum_{j>i} \, [ \, J_{ij}(a_k,a_l) ( \sum_{\VEC{\sigma}} P(\VEC{\sigma}) \delta_{\sigma_i a_k} \delta_{\sigma_j a_l} - P_{ij}(a_k, a_l) )
	\, ]
	\, ]
	=
	\frac{1}{Z} e^{-H_{\script{Potts}}(\VEC{\sigma}|h,J) } 
	\label{eq: max_entropy_distr}
\end{eqnarray}
where $\lambda$, $h_i(a_k)$, and $J_{ij}(a_k,a_l)$ are
Lagrange multipliers, and a Hamiltonian $H_{\script{Potts}}$,
which is called that of the Potts model for $q>2$ (or the Ising model for $q=2$),
and a partition function
$Z$ are defined as
\begin{eqnarray}
	- H_{\script{Potts}}(\VEC{\sigma}|h,J) 
	&=& \sum_i h_i(\sigma_i) + \sum_{i<j} J_{ij}(\sigma_i, \sigma_j)    
	\label{eq: H_for_Potts}
  \hspace*{1em} , \hspace*{1em} Z = 
	\sum_{\VEC{\sigma}} e^{-H_{\script{Potts}}(\VEC{\sigma}|h,J) }
	\label{eq: Z_for_Potts}
\end{eqnarray}
where $h_i(a_k)$ and $J_{ij}(a_k,a_l)$ are interaction potentials 
called fields and couplings.

Although pairwise frequencies $P_{ij}(a_k, a_l)$ reflect
not only direct but indirect correlations in amino acid covariations between sites,
couplings $J_{ij}(a_k,a_l)$ reflect causative correlations only.
Thus, it is essential to estimate fields and couplings from marginal probabilities.
This model is called the inverse Potts model.

\subsubsection{Log-likelihood and log-posterior-probability}
\index{contact prediction!maximum entropy model!log-likelihood}
\index{contact prediction!maximum entropy model!posterior}
\index{contact prediction!maximum entropy model!cross entropy}

Log-posterior-probability and log-likelihood for the Potts model are
\begin{eqnarray}
  \log P_{\script{post}}(h, J|\{\VEC{\sigma}\}) &\propto& \ell_{\script{Potts}}(\{P_i\},\{P_{ij}\} |h, J) + \log P_0(h, J)
	\\
  \ell_{\script{Potts}}(\{P_i\},\{P_{ij}\} |h, J) 
	&=& B \sum_{\VEC{\sigma}} P_{\script{obs}}(\VEC{\sigma}) \log P_{\script{ME}}(\VEC{\sigma}|h, J)
	\label{eq: log-likelihood}
\end{eqnarray}
where 
$P_{\script{obs}} (\equiv \sum_{\tau=1}^{B} \delta_{\VEC{\sigma}\VEC{\sigma}^\tau} / B)$ 
is the observed distribution of $\VEC{\sigma}$ specified with
$\{ P_i(a_k)\}$ and $\{P_{ij}(a_k, a_l)\}$, 
and $B$ is the number of instances;
sequences $\VEC{\sigma}^\tau$ are assumed here to be independently
and identically distributed samples in sequence space.
$P_0(h, J)$ is a prior probability of $(h, J)$. 

Let us define cross entropy\CITE{CM:12} as
the negative log-posterior-probability per instance.
\begin{eqnarray}
S_{0}(h, J | \{P_i\}, \{P_{ij}\} ) 
	&\propto& - ( \log P_{\script{post}}(h, J|\{\VEC{\sigma}\}) ) / B
	\nonumber
	\\
	&\equiv& S_{\script{Potts}}(h, J | \{P_i\}, \{P_{ij}\} )
 	+ R(h, J)
	\label{eq: cross-entropy}
\end{eqnarray}
where the cross entropy $S_{\script{Potts}}$, 
which is the negative log-likelihood per instance for the Potts model, 
and the negative log-prior per instance $R$ are defined as follows.
\begin{eqnarray}
&& S_{\script{Potts}}(h, J | \{P_i\}, \{P_{ij}\}) 
	\equiv - \ell_{\script{Potts}}(\{P_i\},\{P_{ij}\} |h, J) / B
	\\
&&\hspace*{1em}	= \log Z(h, J) - \sum_i \sum_k h_i(a_k) P_i(a_k) 
	- \sum_i \sum_{k} \sum_{j> i} \sum_{l} J_{ij}(a_k, a_l) P_{ij}(a_k,a_l)
	\label{eq: Potts_cross_entropy}
	\\
\lefteqn{
R(h, J) 
	\equiv - \log (P_0(h, J)) / B 
} 
	 \label{eq: log_prior}
	 \label{eq: regularization_term}
\end{eqnarray}
The maximum likelihood estimates of $h$ and $J$, which minimize
the cross entropy with $R=0$, satisfy the following equations.
\begin{eqnarray}
  \frac{\partial \log Z(h, J) }{\partial h_i(a_k) } &=& P_i(a_k)
	\hspace*{1em} \text{ , } \hspace*{1em}
  \frac{\partial \log Z(h, J) }{\partial J_{ij}(a_k, a_l) }
        = P_{ij}(a_k, a_l)
	\label{eq: gradient_of_Z}
\end{eqnarray}
It is, however, hardly tractable to computationally evaluate 
the partition function $Z(h, J)$ for any reasonable system size 
as a function of $h$ and $J$.
Thus, approximate maximization of the log-likelihood
or minimization of the cross entropy is needed to estimate $h$ and $J$.

The minimum of the cross entropy 
with $R=0$
for the Potts model
is just the Legendre transform of $\log Z(h, J)$ 
from $(h, J)$ to $(\{P_i\}, \{P_{ij}\})$, (\Eq{\ref{eq: Potts_cross_entropy}}),
and 
is equal to the entropy of the Potts model
satisfying \Eqs{\Ref{eq: constraints_single_site_marginals} and \Ref{eq: constraints_two_site_marginals}}; 
\begin{eqnarray}
&& S_{\script{Potts}}(\{P_i\}, \{P_{ij}\}) \equiv \min_{h,J} S_{\script{Potts}}(h, J | \{P_i\}, \{P_{ij}\})
	= \sum_{\VEC{\sigma}} - P(\VEC{\sigma}) \log P(\VEC{\sigma})
	\label{eq: Potts_minimum_cross_entropy}
\end{eqnarray}
The cross entropy $S_{\script{Potts}}(h, J | \{P_i\}, \{P_{ij}\})$ in \Eq{\ref{eq: Potts_cross_entropy}}
is invariant under a certain transformation of fields and couplings, 
$J_{ij}(a_k,a_l) \rightarrow J_{ij}(a_k,a_l) - J^1_{ij}(a_k) - J^1_{ji}(a_l) + J^0_{ij}$,
$h_i(a_k) \rightarrow h_i(a_k) - h^0_i + \sum_{j\neq i} J^1_{ij}(a_k)$ for any $J^1_{ij}(a_k)$, $J^0_{ij}$ and $h^0_i$.
This gauge-invariance reduces the number of independent variables 
in the Potts model
to $(q-1)L$ fields and $(q-1)L \times (q-1)L$ couplings.

A prior $P_0(h, J)$ 
yields regularization terms for $h$ and $J$\CITE{CM:12}. 
If a Gaussian distribution is employed for the prior, 
then it will yield $\ell_2$ norm regularization terms. 
$\ell_1$ norm regularization corresponds to the case of exponential priors.
Given marginal probabilities, 
the estimates of fields and couplings are those minimizing the cross entropy.
\begin{eqnarray}
(h, J) &=& \arg \min_{(h, J)} S_{0}(h, J | \{P_i\}, \{P_{ij}\} )
	\,\, , \,\,
S_{0}(\{P_i\}, \{P_{ij}\} ) \equiv \min_{(h, J)} S_{0}(h, J | \{P_i\}, \{P_{ij}\} )
	\label{eq: optimum_h_and_J}
	\label{eq: minimum_cross_entropy}
\end{eqnarray}
Since $S_0(\{P_i\}, \{P_{ij}\} )$ 
is the Legendre transform of $(\log Z(h, j) + R(h, J))$ from $(h, J)$
to $(\{P_i\}, \{P_{ij}\})$,
these optimum $h$ and $J$ can also be calculated from
\begin{eqnarray}
h_i(a_k) &=& - \frac{\partial S_{0}(\{P_i\}, \{P_{ij}\} ) }{\partial P_i(a_k)}
 	\hspace*{1em} , \hspace*{1em}
J_{ij}(a_k,a_l) = -
		\frac{\partial S_{0}(\{P_i\}, \{P_{ij}\} ) }{\partial P_{ij}(a_k,a_l)}
	\label{eq: optimum_h_and_J_from_minimum_S_0}
\end{eqnarray}

In most methods for contact prediction, 
residue pairs are predicted as contacts in the decreasing order of
score ($\mathcal{S}_{ij}$) calculated from fields $\{J_{ij}(a_k,a_l) | 1\leq k, l < q\}$;
\LongVersion{
see \Eqs{\Ref{eq: DI_score} and \Ref{eq: CFN_score}}.
} 
\ShortVersion{
see \Eq{\ref{eq: CFN_score}}.
} 

\subsubsection{Inverse Potts model}
\index{contact prediction!maximum entropy model!inverse Potts model}
\index{contact prediction!maximum entropy model!direct coupling analysis}
\index{inverse Potts model for contact prediction|see{contact prediction!maximum entropy model!inverse Potts model}}
\index{direct coupling analysis for contact prediction|see{contact prediction!maximum entropy model!direct coupling analysis}}

The problem of inferring interactions
from observations of instances has been studied as inverse statistical mechanics, 
particularly inverse Potts model for \Eq{\ref{eq: max_entropy_distr}}, 
in the filed of statistical physics,
as a Markov random field, Markov network or undirected graphical model
in the domain of physics, statistics and information science, 
and as Boltzmann machine in the field of machine learning.

The maximum-entropy approach to 
the prediction of residue-residue contacts
toward protein structure prediction
from residue covariation patterns was
first described in 2002 by Lapedes and collaborators\CITE{GHL:99,LGLS:99,LGJ:02,LGJ:12}.
They estimated 
\LongVersion{
conditional mutual information (CMI) 
defined in \Eq{\ref{eq: conditional_mutual_information}},
} 
\ShortVersion{
conditional mutual information (CMI), 
} 
which was employed as a score
for residue-residue contacts, for each site pair by 
Boltzmann leaning with 
Monte Carlo importance sampling
to calculate equilibrium averages and gradient descent 
to minimize the cross entropy
and successfully predicted contacts for 11 small proteins.

Calculating marginal probabilities for given fields and couplings by
Monte Carlo simulations in Boltzmann machine is very computationally intensive. 
To reduce a computational load, the message passing algorithm,
which is exact for a tree topology of couplings but approximate for the present model, 
is employed instead in mpDCA\CITE{WWSHH:09}.
Because even the message passing algorithm
is too slow to be applied to
a large-scale analysis across many protein families,
the mean field approximation
is employed 
in mfDCA\CITE{MPLBMSZOHW:11,MCSHPZS:11}; 
\LongVersion{
see \Eq{\ref{eq: J_MF}};		
} 
$J^{MF} = - C^{-1}$,
where $C_{ij}(a_k,a_l) \equiv P_{ij}(a_k,a_l) - P_i(a_k) P_j(a_l)$. 	
In the mean field approximation, 
a bottleneck in computation
is the calculation of the inverse of a covariance matrix $C$
\LongVersion{
that is a $(q-1) L \times (q-1) L$ matrix 
defined in \Eq{\ref{eq: covariance_matrix}}.
} 
\ShortVersion{
that is a $(q-1) L \times (q-1) L$ matrix. 
} 
In the mean field approximation,
a prior distribution in \Eq{\ref{eq: log_prior}} is ignored and pseudocount is
employed instead of regularization terms to make the covariance matrix invertible.

The Gaussian approximation (a continuous multivariate Gaussian model) 
for the probability distribution of 
sequences is employed together 
with an exponential prior (an $\ell_1$ regularization term) in PSICOV\CITE{JBCP:12}, 
and with a normal-inverse-Wishart (NIW) prior,
which is a conjugate distribution of the multivariate Gaussian,
in GaussDCA\CITE{BZFPZWP:14}. 
The use of NIW prior has a merit that
fields and couplings can be analytically formulated; see \Eqs{\Ref{eq: J_NIW} and \Ref{eq: h_NIW}}.

All methods based on the Gaussian approximation   
employ the analytical formula 
for couplings, $J \simeq - C^{-1} = - \Theta$, 
which are essentially as same as the mean field
approximation with a difference that 
the covariance matrix ($C$) or precision matrix ($\Theta$) is differently estimated based on the various priors.
The mean field and Gaussian approximations may be appropriate to systems of dense and weak couplings but
questionable for sparse and strong couplings that is the characteristic of 
residue-residue
contact networks.
Although the mean field
and Gaussian approximations
successfully predict residue-residue contacts in proteins,
it has been shown\CITE{BLCC:16,CFFMW:17} that they do not give 
the accurate estimates of fields and couplings in proteins.

A pseudo-likelihood with Gaussian priors ($\ell_2$ regularization terms) 
is maximized to estimate fields and couplings in plmDCA\CITE{ELLWA:13,EHA:14}
for the Potts model with sparse interactions as well as reducing computational time; 
see \Eq{\ref{eq: symmetric pseudo-likelihood maximization}} 
for the symmetric plmDCA
and
\Eq{\ref{eq: asymmetric pseudo-likelihood maximization}}
for the asymmetric plmDCA. 
The asymmetric plmDCA method\CITE{EHA:14} requires less computational time 
and fits particularly with parallel computing.

GREMLIN\CITE{KOB:13} employs together with pseudo-likelihood 
Gaussian priors
that depend on site pair, although
its earlier version\CITE{BKCLL:11} employed $\ell_1$ regularizers,
which may be more appropriate to systems of sparse couplings.
The $\ell_1$ regularizers appear to learn parameters that are closer to their true strength,
but the $\ell_2$ regularizers appear to be as good as the $\ell_1$ regularizers for 
the task of contact prediction that requires the relative ranking of the interactions 
and not their actual values\CITE{KOB:13}.

One of approaches to surpass 
the pseudo-likelihood approximation
for systems of sparse couplings
may be the adaptive cluster expansion (ACE) of cross entropy\CITE{CM:11,CM:12,BLCC:16}, in which
cross entropy is approximately minimized by taking account of 
only site clusters the incremental entropy (cluster entropy) 
of which by adding one more site
is significant.  
In this method\CITE{BLCC:16}, a Boltzmann machine is employed to refine
fields and couplings and also to calculate 
model correlations 
such as single-site and pairwise amino acid frequencies under given fields and couplings. 
The results of the Boltzmann machine for both biological and artificial models 
showed that ACE outperforms plmDCA in recovering single-site marginals 
(amino acid frequencies at each site) and
the distribution of the total dimensionless energies (${H}_{\script{Potts}}(\VEC{\sigma})$)
\CITE{BLCC:16};
those models were
a lattice protein, trypsin inhibitor, HIV p7 nucleocapsid protein,
multi-electrode recording of cortical neurons, and Potts models on Erid\"{o}s-R\'{e}nyi random graphs.
More importantly ACE 
could accurately recover the true fields $h$ and couplings $J$ corresponding to
Potts states with $P_i(a_k) \geq 0.05$ for 
Potts models ($L=50$) on Erid\"{o}s-R\'{e}nyi random graphs\CITE{BLCC:16}.
On the other hand, plmDCA gave accurate estimates of couplings at weak regularization  
for well sampled single-site probabilities, but less accurate fields.
Also, plmDCA yielded less well inferred fields and couplings
for single-site and two-site probabilities not well sampled, indicating that 
not well populated states should be merged.
As a result, the distribution of the total energies\CITE{BLCC:16} and
the distribution of mutations with respect to the consensus sequence were 
not well reproduced\CITE{CFFMW:17}.
Similarly,
the mean field approximation could not reproduce two-site marginals and even single-site marginals
\CITE{CFFMW:17}
and the Gaussian approximation could not well reproduce 
the distribution of mutations with respect to the consensus sequence\CITE{BLCC:16}.

However, the less reproducibility of couplings does not necessarily indicate 
the less predictability of residue-residue contacts,
probably because in contact prediction the relative ranking of 
\LongVersion{
scores (\Eqs{\Ref{eq: DI_score} or \Ref{eq: CFN_score}}) 
} 
\ShortVersion{
scores (\Eq{\ref{eq: CFN_score}}) 
} 
based on couplings is more important than their actual values.
ACE with the optimum regularization strength with respect to the reproducibility of fields and
couplings showed less accurate contact prediction than plmDCA and mfDCA.
For ACE to show comparable performance of contact prediction with plmDCA,
regularization strength had to be increased from $\gamma = 2/B = 10^{-3}$ to $\gamma = 1$ 
for Trypsin inhibitor, making couplings strongly damped and then 
the generative properties of inferred models lost\CITE{BLCC:16}.

\begin{table}[!ht]
\caption{
\label{tbl: software}
Free softwares/servers for the direct coupling analysis. 
} 

\vspace*{1em}
\footnotesize
\begin{tabular}{lll}
\hline
\\
Name	& Methods	& URL		\\
\hline	
EVcouplings \CITE{MCSHPZS:11}		& mfDCA	& http://evfold.org	\\
EVcouplings, plmc\CITE{TPIHBSM:16,WRIGSM:16}	
			& mf/plmDCA & https://github.com/debbiemarkslab \\
DCA \CITE{MPLBMSZOHW:11,MCSHPZS:11}	& mfDCA	& http://dca.rice.edu/portal/dca/home
				\\
GaussDCA \CITE{BZFPZWP:14} & GaussDCA
       & http://areeweb.polito.it/ricerca/cmp/code \\
FreeContact \CITE{KHKMR:14}    & mfDCA, PSICONV	
			& http://rostlab.org/owiki/index.php/FreeContact
			\\
plmDCA \CITE{ELLWA:13,EHA:14}	& plmDCA & http://plmdca.csc.kth.se/	\\
		&	& https://github.com/pagnani/plmDCA	\\
CCMpred \CITE{SGS:14}  & plmDCA	& performance-optimized software \\
			&	& https://github.com/soedinglab/ccmpred	\\
GREMLIN	\CITE{BKCLL:11,KOB:13}	& GREMLIN & http://gremlin.bakerlab.org/	\\
ACE \CITE{CM:11,CM:12,BLCC:16}
        & ACE   & https://github.com/johnbarton/ACE \\
persistent-vi\CITE{IM:16} 	& Persistent VI & https://github.com/debbiemarkslab \\
\hline	
\end{tabular}

\end{table}
\subsection{Partial correlation of amino acid cosubstitutions between sites at each branch of a phylogenetic tree}
\index{contact prediction!partial correlation of amino acid cosubstitutions}
\index{partial correlation of amino acid cosubstitutions for contact prediction|see{contact prediction!partial correlation}}

In the DCA analyses on residue covariations 
between sites in a multiple sequence alignment (MSA),
phylogenetic biases, 
which are sequence biases due to phylogenetic relations between species, 
in the MSA 
must be removed 
as well as indirect correlations between sites,
but instead are reduced 
by taking weighted averages over homologous sequences
in the calculation of single and pairwise 
\LongVersion{
frequencies of amino acids; see \Eq{\ref{eq: phylogenetic_bias}}.
} 
\ShortVersion{
frequencies of amino acids.
} 

Needless to say, it is supposed that 
observed patterns of covariation were caused by molecular coevolution
between sites. 
Whatever caused covariations found in the MSA,
it has been confirmed that
they can be utilized to predict residue pairs in close proximity 
in a three dimensional structure.
Talavera et al.\CITE{TLW:15} claimed, however, 
that covarying substitutions were mostly found 
on different branches of the phylogenetic tree, indicating that they 
might or might not be attributable to coevolution.

In order to remove phylogenetic biases and also to respond to such a claim above,
it is meaningful to study covarying substitutions between sites
in a phylogenetic tree-dependent manner. 
Such an alternative approach was taken
to infer coevolving site pairs from direct correlations between sites 
in concurrent and compensatory substitutions 
within the same branches of a phylogenetic tree\CITE{M:13}.
In this method,
substitution probability and mean
changes of physico-chemical properties of side chain accompanied
by amino acid substitutions at each site in each branch of the tree
are  estimated  with  the  likelihood  of  each  substitution  to  detect
concurrent  and  compensatory  substitutions.
Then, 
partial correlation coefficients of
the vectors of their characteristic changes accompanied by substitutions,
substitution probability and mean changes of physico-chemical properties, 
along branches between sites are calculated to extract 
direct correlations in coevolutionary substitutions 
and employed as a score for residue-residue contact. 
The accuracy of contact prediction by this method 
was comparable with
that by mfDCA\CITE{M:13}.
This method, however, has a drawback to be computationally intensive,
because an optimum phylogenetic tree must be estimated.

\section{Machine learning methods to augment the contact prediction accuracy based on amino acid coevolution}
\index{contact prediction!machine learning}
\index{machine learning for contact prediction|see{contact prediction!machine learning}}
\index{contact prediction!deep learning|see{contact prediction!machine learning}}
\index{deep learning for contact prediction|see{contact prediction!machine learning}}
\index{contact prediction!deep neural network|see{contact prediction!machine learning}}
\index{deep neural network for contact prediction|see{contact prediction!machine learning}}

\begin{table}[!ht]
\caption{
\label{tbl: post-processing}
Machine learning methods that combine predicted direct couplings
with other sequence/structure information.
} 

\vspace*{1em}
\footnotesize
\begin{tabular}{lll}
\hline
\\
Name	&	Basic method	& Post-processing	\\
\hline	
PconsC3 \CITE{SMHEE:16}	& plmDCA, GaussDCA 	& 5 layer DNN; http://c3.pcons.net.
					PconsC\CITE{SAE:13},C2\CITE{SAE:14} 	\\
MetaPSICOV	& PSICOV,mfDCA, & A two stage neural network predictor; CONSIP2 pipeline	\\
 \hspace*{1em}\CITE{KJ:14,JSKT:15,KJ:16}		
	& \hspace*{1em}GREMLIN/CCMpred	
			& http://bioinf.cs.ucl.ac.uk/MetaPSICOV	\\
RaptorX \CITE{WSLZX:17}	& CCMpred & Ultra-deep learning model consisting of 1- and
		\\
	& &
	2-dimensional convolutional residual neural networks
		\\
	& &
	http://raptorx.uchicago.edu/ContactMap/
		\\
iFold \CITE{CASP12}	& 	& Deep neural network (DNN)
	\\
EPSILON-CP & PSICOV, GREMLIN, & 4 hidden layer neural network
					\\
\multicolumn{2}{r}{mfDCA,CCMpred,GaussDCA} & \hspace*{1em} with 400-200-200-50 neurons\CITE{SSB:17}
 	\\
\hline	
\end{tabular}

\end{table}
All the DCA methods such as mfDCA, plmDCA, GREMLIN, and PSICOV 
predict significantly nonoverlapping sets of contacts\CITE{JSKT:15,KJ:16,WZPY:16}.
Then, increasing prediction accuracy by combining 
their predictions together with other sequence/structure information  
have been attempted\CITE{SAE:13,SAE:14,SMHEE:16,KJ:14,JSKT:15,KJ:16,WSLZX:17,SSB:17};
see \Table{\ref{tbl: post-processing}}.

PconsC\CITE{SAE:13} combines the predictions of PSICOV and plmDCA
into a machine learning method, random forests,
and employs alignments with HHblits\CITE{RBHS:12} and jackHMMer\CITE{JEP:10} 
at four different e-value cut-offs.
Five-layer neural network
is employed instead of random forests 
in PconsC2\CITE{SAE:14}, and plmDCA and GaussDCA are 
employed
in PconsC3\CITE{SMHEE:16}. 
A receptive field consisting of $11 \times 11$ predicted contacts around each residue
pair is taken into account in each layer except the first one.

MetaPSICOV\CITE{JSKT:15,KJ:16} combines the predictions of PSICOV, mfDCA, and CCMpred/GREMLIN
into the first stage of a two-stage neural network predictor
together with a well-established ``classic'' machine learning contact
predictor, which utilizes
many features such as
amino acid profiles, predicted
secondary structure and solvent accessibility along
with sequence separation predicted, as an additional source of information
for a little depth of MSAs.
The second stage analyses the output of the first stage 
to eliminate outliers and to fill in the gaps in the contact map.
On a set of 40 target domains with a median family size of
around 40 effective sequences in CASP11, 
CONSIP2 server achieved an average top-$L/5$ long-range contact precision of 27\%\CITE{KJ:16}.

Wang et al.\CITE{WSLZX:17} have also shown that
a ultra-deep neural network (RaptorX)
can significantly improve contact prediction based on
amino acid coevolution.
They have modeled
short-range and long-range correlations in
sequential and structural features with respect to
complex sequence-structure relationships in proteins
by one-dimensional and two-dimensional 
deep neural networks (DNN), respectively.
Both the DNNs are convolutional residual neural networks.
The 1D DNN performs convolutional transformations, with respect to residue position, of 
sequential features such as 
position-dependent scoring matrix, 
predicted 3-state secondary structure and 3-state solvent accessibility.
The 2D DNN does 2D convolutional transformations of  
pairwise features such as coevolutional information calculated by CCMpred, 
mutual information, pairwise contact potentials as well as
the output of the 1D DNN converted by a similar operation to outer product.
Residual neural networks are employed 
because they can pass both linear and nonlinear informations
from initial input to final output,
 making their training relatively easy.
 
\section{Performance of contact prediction}
\index{contact prediction!performance}
\index{contact prediction!accuracy}

New statistical methods based on the direct coupling analysis
are confirmed in
various benchmarking studies\CITE{CASP11:16,CASP12,KOB:13,WZPY:16}
to show remarkable accuracy of contact prediction,
although deep, stable alignments are required.
They can more accurately detect a higher number of  contacts
between residues, which are very distant along sequence\CITE{MPLBMSZOHW:11}. 
The top-scoring
residue couplings are not only sufficiently accurate but also well-distributed 
to define the 3D protein fold with remarkable accuracy\CITE{MCSHPZS:11};
this observation was quantified by computing, from sequence alone, 
all-atom 3D structures of fifteen test proteins from different
fold classes, ranging in size from 50 to 260 residues, including a G-protein coupled receptor.
The contact prediction performs relatively better on $\beta$ proteins
than on $\alpha$ proteins\CITE{M:13}.
These initial findings on a limited number of proteins 
were confirmed as a general trend in a large-scale comparative assessment of contact prediction methods
\CITE{WZPY:16,ANBHC:16}. 

In CASP12, RaptorX performed the best in terms of F1 score for top $L/2$ long- and medium-range contacts
of 38 free-modeling (FM) targets; the total F1 score of RaptorX was better by about 7.6 and 10.0 \% than
the second and third best servers, iFold\_1 and the revised MetaPSICOV, respectively\CITE{WSLZX:17,CASP12}.
Tested on 105 CASP11 targets, 76 past CAMEO hard targets, and 398 membrane
proteins, the average top $L$ ($L/10$) long-range prediction accuracies of RaptorX
are 0.47(0.77) in comparison with 0.30(0.59) for MetaPSICOV and 0.21(0.47) for CCMpred\CITE{WSLZX:17,CASP12}.
 
\subsection{MSA dependence of contact prediction accuracy}

\noindent
In the direct-coupling-based methods, the accuracy of predicted contacts 
depends on the depth\CITE{M:13,KOB:13,WZPY:16} 
and quality of multiple sequence alignment (MSA) for a target.
$5 \times L$ (protein length) aligned sequences may be desirable for accurate contact predictions\CITE{KOB:13},
although attempts to improve prediction methods for fewer aligned sequences have been
made\CITE{SAE:13,SAE:14,SMHEE:16,WSLZX:17}. 
PconsC3 can be used for families with as little as 100 effective sequence members\CITE{SMHEE:16}.
Also, RaptorX\CITE{WSLZX:17} attained top-$L/2$-accuracy$>0.3$ 
for long-rang contacts even by using MSAs with 
\LongVersion{
20 effective sequence members ($M_{\script{eff}} \sim 20$);
see \Eq{\ref{eq: weight_for_sequence}} for $M_{\script{eff}}$.
} 
\ShortVersion{
20 effective sequence members.
} 

Deepest MSAs including a target sequence were built 
with various values of E-value cutoff\CITE{SAE:13} and coverage parameters\CITE{JSKT:15,KJ:16}
in sequence search and alignment programs based on the hidden Markov models such as 
HHblits and jackHMMer.
Although prediction performance tends to increase in general as alignment depth is deeper\CITE{M:13},
it was reported\CITE{KJ:16} that
in the case of transmembrane domains, building too deep
alignments could result in unrelated sequences or drifted
domains being included. 
To increase alignment quality, 
E-value and coverage parameters may be carefully 
tuned for each alignment\CITE{KJ:16}.
In the case of alignments that might contain regions of partial matches,
a too stringent sequence coverage requirement could result in
missing related sequences. 
On the other hand, a too permissive sequence coverage requirement
could pick up unrelated sequences, permitting many partial matches.
A trade-off is required between the effective number of sequences and
sequence coverage, and 
an appropriate E-value must be chosen not to much decrease 
both alignment depth and sequence coverage\CITE{HCSRSM:12}. 

\section{Contact-guided de novo protein structure prediction}
\index{contact prediction!structure prediction}
\index{structure prediction based on predicted contacts|see{contact prediction!structure prediction}}

\begin{table}[!ht]
\caption{
\label{tbl: 3D_pred}
Contact-guided de novo protein structure prediction methods and servers. 
} 

\vspace*{1em}
\footnotesize
\begin{tabular}{lll}
\hline
\\
Name	& Contact prediction	& 	\\
\hline	
EVfold \CITE{MCSHPZS:11,MHS:12}
	& mfDCA / plmDCA
	& 
	Using distance geometry algorithm\CITE{HKC:83} and 
	\\
\multicolumn{2}{l}{\hspace*{1em}
/EVfold\_membrane \CITE{HCSRSM:12}
}
		& simulated annealing of CNS\CITE{B:07};
		http://evfold.org/	\\
DCA-fold\CITE{SMWHO:12} & mfDCA
	& Simulated annealing using a coarse-grained
	\\
	& & molecular dynamics for a C$_{\alpha}$ model
	\\
FRAGFOLD
	& MetaPSICOV 
	& Combining fragment-based folding algorithm\CITE{JBCMSSW:05} 
	\\
\hspace*{1em}/FILM3	& & with PSICOV\CITE{KJ:14} and with MetaPSICOV\CITE{JSKT:15}.
		\\
	& & FILM3\CITE{NJ:12} is employed instead of FRAGFOLD\CITE{J:01} 
		\\
	& & for transmembrane proteins.
		\\
CONFOLD \CITE{ABCC:15}	& 
	EVFOLD / FRAGFOLD 
	&
	Two-stage contact-guided de novo protein folding, 
		\\
	& (PSIPRED for 2nd structures) & using distance geometry simulated annealing
		\\
	& & protocol in a revised CNS v1.3.
		\\
	& & http://protein.rnet.missouri.edu/confold/
		\\
Rosetta \CITE{KCB:04}\CITE{OKWLDB:16} & GREMLIN
		& Fragment assembly
	\\
\hline	
\end{tabular}

\end{table}
\noindent
It is a primary obstacle to de novo structure prediction
that current methods and computers cannot make it feasible to
adequately sample 
the vast conformational space
a protein might take in the precess of folding into the native structure\CITE{KBBB:09}.
Thus, it is critical 
whether residue-residue proximities inferred with direct
coupling analysis can provide sufficient information 
to reduce a huge search space for a protein fold, 
without any known 3D structural information of the protein.

Algorithms are needed to fold proteins into native folds based on
contact information; see \Table{\ref{tbl: 3D_pred}}.
Distance geometry generation\CITE{HKC:83,BG:85} of 3D structures,
which may be followed by energy minimization and molecular dynamics,
will be just the primary one.
In EVfold\CITE{MCSHPZS:11},
contacts inferred by direct coupling analysis
and predicted secondary structure information 
are translated into a set of
distance constraints for the use of a distance geometry
algorithm in the Crystallography and NMR System (CNS)\CITE{B:07}. 
It was confirmed that the evolutionary inferred contacts 
can sufficiently reduce 
a search space in the structure predictions 
of 15 test proteins from different fold classes\CITE{MCSHPZS:11}, and
of 11 unknown and 23 known transmembrane protein structures\CITE{HCSRSM:12}.
Because distance constraints from predicted contacts may be 
partial in a protein sequence,	
they should be embedded into \textit{ab initio} structure prediction methods.

Su{\l}kowska et al. also showed that
a simple hybrid method, called DCA-fold, integrating mfDCA-predicted contacts
with an accurate knowledge of secondary structure
is
sufficient to fold proteins in the range 
of 1-3 \AA\  resolution\CITE{SMWHO:12}.
In this study, simulated annealing using a coarse-grained molecular dynamics 
model was employed for a C$_{\alpha}$ chain model, in which
C$_{\alpha}$s 
interact with each other 
with a contact potential 
approximated by a Gaussian function
and a torsional potential depending on C$_{\alpha}$ dihedral angles at each position.

Adhikari et al.\CITE{ABCC:15}  
studied a way to effectively encode secondary structure information
into distance and dihedral angle constrains that complement
long-range contact constraints,
and revised  
the CNS v1.3 to effectively use secondary structure constraints
together with predicted long-range constraints;
CONFOLD\CITE{ABCC:15} consists of two stages.
In the first stage
secondary structure information is converted into distance,
dihedral angle, and hydrogen bond constraints, and then
best models are selected by executing the
distance geometry simulated annealing.
In the second stage self-conflicting contacts 
in the best structure predicted in the first stage 
are removed, 
constrains based on the secondary structures are refined,
and again the distance geometry simulated annealing is executed.

Baker group\CITE{OKWLDB:16} 
embedded contact constraints predicted by GREMLIN\CITE{KOB:13} 
as sigmoidal constraints to overcome noise in the Rosetta\CITE{KCB:04}
conformational sampling and refinement.
They found
that model accuracy will be generally improved, 
if more than 3 L (protein length) sequences are available, 
and that 
large topologically
complex proteins
can be modeled with close to atomic-level accuracy
without knowledge of homologous structures,
if there are enough homologous sequences available. 

On the other hand, 
a fragment-based folding algorithm FRAGFOLD was combined
with PSICOV \CITE{KJ:14} and with MetaPSICOV\CITE{JSKT:15,KJ:16};
In this approach, predicted contacts are converted into 
additional energy terms for FRAGFOLD in addition
to the pairwise potentials of mean force and solvation\CITE{JSKT:15,KJ:16}.
FILM3\CITE{NJ:12}, 
with constraints based on predicted contacts
and ones approximating Z-coordinate values within the lipid membrane,
is
employed instead of FRAGFOLD for transmembrane proteins.

RaptorX\CITE{WSLZX:17} employed the CNS suite\CITE{B:07} 
to generate 3D models from predicted contacts and 
secondary structure 
converted to distance, angle and h-bond restraints,
and could yield TMscore $> 0.6$ for 203 of 579 test proteins,
while using MetaPSICOV and CCMpred could do so
for 79 and 62, respectively.

\subsection{How many predicted contacts should be used to build 3D models?}

\noindent
The number of feasible contacts surrounding a residue in a protein
is about 6.3 
\CITE{MJ:96}, 
which corresponds to 
the maximum number of contacts per a protein, $6.3 L/2$, where $L$
denotes protein length.  
However,
more than 50\% of known 3D structures in the PDB 
have less than $2 L$ contacts,
and in the test on 15 proteins in
EVfold benchmark set, less than $1.6 L$ predicted contacts
yielded best results\CITE{ABCC:15}.
In the original EVfold, the optimal number of
evolutionary constraints was in the order of $0.5 L$ to $0.7 L$\CITE{HCSRSM:12}.
Because 
prediction accuracy tends to decrease as the rank of contact score increases, and
different proteins need different numbers of predicted contacts 
to be folded well, 
protein folds were generated with a wide range of the number of predicted contacts,
and then best folds were selected; 
from $30$ to $L$ in EVfold\CITE{HCSRSM:12}, and
from $0.4 L$ to $2.2 L$ in CONFOLD\CITE{ABCC:15}.
In RaptorX, the top $2L$ predicted contacts irrespective of site separation 
were converted to distance restraints\CITE{WSLZX:17}.
On the other hand, Jones group reported\CITE{KJ:14}
that artificially truncating the list of predicted contacts
was likely to remove useful information 
to fold a protein with FRAGFOLD and PSICOV, in which
the weight of a given predicted contact 
is determined 
\LongVersion{
by its positive predictive value; 
see section \ref{sec: L11_norm}.
} 
\ShortVersion{
by its positive predictive value.
} 

\section{Evolutionary direct couplings between residues not contacting in a protein 3D structure}

Needless to say, 
evolutionary constraints do not only originate in 
intra-molecular contacts
but also result from inter-molecular contacts/interactions.
Even in the case of intra-molecular contacts, 
if there are structural variations 
including ones due to conformational changes in a protein family,
evolutionary constraints
will reflect the alternative conformations\CITE{MPLBMSZOHW:11,HCSRSM:12,AOKB:17}.
Also, intra-molecular residue couplings
may contain useful 
information of ligand-mediated residue couplings
\CITE{MPLBMSZOHW:11,OKWLDB:16}.
On the other hand, 
inter-molecular contacts may allow us to
predict protein complexes, and 
are useful to build protein-protein interaction networks at a residue level.
 
\subsection{Structural variation including conformational changes}
\index{contact prediction!structural variation}

\noindent
MSA contains information on all members of the protein family,
and direct couplings between residues estimated from the MSA 
reflect the structures of all members.
It was shown\CITE{AOKB:17} that
74 \% of top $L/2$ direct couplings residue pairs
that are more than 5 \AA\ apart in the target structures of
3883 proteins
are less than 
5 \AA\ apart in at least one homolog structure. 

Conformational change is an interesting case of 
structural variation. 
Many proteins adopt different conformations as part
of their functions\CITE{TT:09}, indicating that
protein flexibility is as important as structure on
biological function.
Protein flexibility around the energy minimum 
can be studied by sampling around the native structure
in normal mode/principal component analysis,
coarse-grained elastic network model, and
short-timescale MD simulations.
However, distant conformers that require large
conformational transitions are difficult to predict.
If conformational changes are essential on
protein functions, evolutionary constraints
will reflect the multiple conformations.
Toth-Petroczy et al.\CITE{TPIHBSM:16} 
showed that coevolutionary information may reveal 
alternative structural states of disorderd regions.

Morcos et al.\CITE{MPLBMSZOHW:11} found that
some of top predicted contacts in the response-regulator DNA-binding domain family
(GerE, PF00196) conflict
with the structure (PDB ID 3C3W) of the full-length response-regulator DosR of M. tuberculosis,
but are compatible with the structure (PDB ID 1JE8) of DNA-binding domain of 
\textit{E. coli} NarL.

Sutto et al.\CITE{SMVG:15}
combined coevolutionary data and molecular dynamics simulations 
to study protein conformational heterogeneity;
the Boltzmann-learning algorithm with $\ell_2$ regularization terms 
was employed to extract direct couplings 
between sites in homologous protein sequences, and 
a set of conformations consistent with the observed residue couplings
were generated by exhaustive sampling simulations 
based on a coarse-grained protein model. 
Although the most representative structure was consistent with the 
experimental fold, the various regions of the sequence showed
different stability, indicating conformational changes\CITE{SMVG:15}.

Sfriso et al.\CITE{SDMEAO:16}
made an automated pipeline
based on discrete molecular dynamics guided by predicted contacts
for the systematic identification of functional
conformations in proteins, and identified
alternative conformers 
in 70 of 92 proteins
in a validation set of proteins in PDB;
various conformational transitions 
are
relevant to those conformers, such as
open-closed, rotation, rotation-closed, 
concerted, and miscellanea of complex motions.

\subsection{Homo-oligomer contacts}
\index{contact prediction!homo-oligomer contacts}

Intra-molecular contacts that conflict with the native fold
may indicate homo-oligomer contacts\CITE{AOKB:17}.
Such a case was confirmed 
for homo-oligomer contacts 
in the ATPase domain of 
nitrogen regulatory protein C-like  
sigma-54 dependent transcriptional activators\CITE{MPLBMSZOHW:11}
and between transmembrane helices\CITE{HCSRSM:12}.
It was pointed out\CITE{HCSRSM:12} that
the identification of evolutionary couplings due to homo-oligomerization
is not only meaningful in itself but also
useful because 
their removal improves the accuracy of the structure prediction for the monomer.

\subsection{Residue couplings mediated by binding to a third agent}
\index{contact prediction!residue couplings mediated by binding to a third agent}

Direct couplings between residues found by the DCA analysis
can be mediated\CITE{MPLBMSZOHW:11} by their interactions with a third agent,
i.e., ligands, substrates, RNA, DNA, and other metabolites.
This indicates that binding sites with such a agent may be found as
residue sites directly coupled but not in contact.

If interactions with a third agent requires too specific residue type at a certain site,
then the residue type will be well conserved at the binding sites. This often occurs, and
has been utilized to identify binding sites. However, 
the interactions for binding are less specific but certainly restricted,
direct couplings between residues around the binding sites may occurs.

Hopf et al.\CITE{HCSRSM:12} devised a total evolutionary coupling score, 
which is defined as EC values summed over all high-ranking pairs 
involving a given residue and normalized by their average over all high-ranking pairs,
and showed that residues with high total coupling scores line
substrate-binding sites and affect signaling or transport 
in transmembrane proteins, Adrb2 and Opsd. 

\section{Heterogeneous protein-protein contacts}
\index{contact prediction!heterogeneous protein-protein contacts}

An application of the direct coupling analysis to
predict the structures of protein complexes is 
straightforward.
In place of a MSA of a single protein family, 
a single MSA that is built by concatenating
the multiple MSAs of multiple protein families every species 
can be employed to extract 
direct couplings between sites of different proteins by removing 
indirect intra- and inter-protein couplings
\CITE{PHAV:97,SPSLAGL:08,WWSHH:09,HCSRSM:12}.

A critical requirement for sequences to be concatenated is, however, that 
respective sets of the protein sequences
must have the same evolutionary history to coevolve. 
In other words, phylogenetic trees built from the respective sets of sequences 
employed for the protein families must have at least the same topology.
One way to build a set of cognate pairs of protein sequences  
is to employ orthologous sequences for each protein family, the phylogenetic tree of which
coincides with that of species.
Thus, 
a genome-wide analysis of finding protein-protein interactions 
based on protein sequences is not so simple.

Weigt et al.\CITE{WWSHH:09} successfully applied 
the direct coupling analysis to the bacterial two-component signal transduction system
consisting of sensor kinase (SK) and response regulator (RR),
which are believed\CITE{SPSLAGL:08} to interact specifically with each other in most cases
and often revealed by adjacency in chromosomal location.
This analysis is based on the fact that
in prokaryotes cognate pairs are often encoded in the same operon.
Genome-sequencing projects have revealed that most organisms
contain large expansions of a relatively small number of signaling
families\CITE{SPSLAGL:08}. 
However, it is not as simple as in prokaryotes to 
build a set of cognate pairs of those protein sequences in eukaryotes.

Hopf et al.\CITE{HSRGKSBM:14} 
developed a contact score, EVcomplex, 
for every inter-protein residue pair based on the overall inter-protein
EC score distributions,
evaluated its performance
in blinded tests on 76 complexes of known 3D structure, 
predicted protein-protein contacts in 32 complexes of unknown structure, 
and then demonstrated how evolutionary direct couplings 
can
be used 
to distinguish between interacting and non-interacting
protein pairs 
\LongVersion{
in a large complex; see section \ref{sec: DI} for EC score.
} 
\ShortVersion{
in a large complex.
} 
In their analysis, protein sequence pairs that are encoded close on \textit{E. coli} genome
were employed to reduce incorrect protein pairings.

\section{Discussion}

Determination of protein structure is essential to understand
protein function.
However,
despite significant effort to explore unknown folds in
the protein structural space,
protein structures determined by experiment
are far less than known protein families.
Only about 41--42\% of the Pfam families\CITE{FCEEMMPPQSSTB:16}
(Pfam-A release 31.0, 16712 families)
include at least one member whose structure is known.
The number and also the size  of protein families will further grow
as genome/metagenome sequencing projects proceed with next-generation
sequencing technologies.
Thus, accurate de novo prediction of three-dimensional structure
is desirable
to catch up with the high growing speed of protein families with unknown folds.
Coevolutionary information can be used to predict not only proteins but also RNAs\CITE{WRIGSM:16} 
and those complexes,
together with experimental informations such as X-ray, NMR, SAS, FRET, crosslinking, Cryo-EM, and others.

Here, statistical methods for disentangling direct from indirect 
couplings between sites with respect to evolutionary variations/substitutions
of amino acids in homologous proteins have been briefly reviewed.
Dramatic improvements on contact prediction
and successful 3D de novo predictions based on predicted contacts
are described in details in
the recent reports of CASP-11\CITE{CASP11:16} and CASP-12 meetings\CITE{CASP12}.
Machine learning methods, particularly deep neural network (DNN) such as MetaPSICOV, iFold, and RaptorX, 
have shown to significantly augment 
contact prediction accuracy 
based on coevolutionary information.
However, the present state-of-the-art DNN methods are, at least at the very moment, not powerful enough to extract
coevolutionary information directly from homologous sequences.
It was reported that without coevolutionary strength produced by CCMpred 
the top $L/10$ long-range prediction accuracy of RaptorX might drop by 0.15 for soluble proteins and
more for membrane proteins\CITE{WSLZX:17},
indicating that the direct coupling analysis is still essential for contact prediction.

The primary requirement for the direct coupling analysis
is a high quality deep alignment. 
However, 
genome/metagenome sequencing projects 
provide more genetic variations
from which more accurate and more comprehensive information on
evolutionary constraints can be extracted.
One of problems is that species being sequenced may be strongly biased to
prokaryotes, making it hard to analyze eukaryotic proteins 
based on coevolutionary substitutions.
Experiments of vitro evolution may be useful to provide sequence variations 
for eukaryotic proteins\CITE{OKWLDB:16}. 

For a large-scale of protein structure prediction,
computationally intensive methods such as the ACE and Boltzmann machine (MCMC and mpDCA)
can hardly be employed.  The Gaussian approximation with a normal-inverse-Wishart prior,
the Gaussian approximations with other priors (PSICOV) and mean field approximation (mfDCA)
are fast enough but 
their performance of contact prediction tends to be compared unfavorably with the pseudo-likelihood
approximation (plmDCA), indicating that they may be inappropriate for proteins with sparse couplings.

The accurate estimates of fields and couplings are very informative
in evaluating the effects ($\Delta {H}_{\script{Potts}}$) of mutations\CITE{HIPSSSM:17},
identifying protein family members and also 
studying folding mechanisms\CITE{MSCOW:14,JGSCM:16}
and protein evolution\CITE{M:17}.
It should be
also examined whether the distribution of dimensionless energies (${H}_{\script{Potts}}$) over homologous proteins
can be well reproduced.  
Accuracy of estimates of fields and couplings 
and the distribution of dimensionless energies depends on 
regularization parameters or the ratio of pseudocount\CITE{BLCC:16,M:17},
and therefore they should be optimized. 
It was also pointed out that group $L_1$ regularization performs better than $L_2$
for the maximum pseudolikelihood method\CITE{IM:16}.
The ACE algorithm, which can be applied only for systems of sparse couplings, 
may be more favorable
with respect to computational load 
for the estimation of fields and couplings 
than Boltzmann learning with Monte Carlo simulation or with message passing.
However, both the methods are computationally intensive.
Recently, another approach consisting of two methods named persistent-vi and Fadeout, in which
the posterior probability density with horseshoe prior 
is approximately estimated
by using variational inference and
noncentered parameterization for such a sparsity-inducing prior, 
has shown to perform better with twofold cpu time than 
the maximum pseudolikelihood method with $L_2$ and group $L_1$ regularizations\CITE{IM:16}.

The remarkable advances of sequencing technologies and also statistical methods 
are likely to bring many targets within range of the present approach in the near future, 
and have a potential to transform the field\CITE{CASP11:16}.

\noindent
\large
\section*{Appendix}
\addcontentsline{toc}{section}{Appendix}
\normalsize

\setcounter{section}{9}

\small

\ShortVersion{
\noindent
An appendix described in full will be found in the article\CITE{M:17b} submitted to the arXiv.
} 

\LongVersion{
\subsection{Direct Coupling Analysis for amino acid covariations between sites}

\subsubsection{A reweighted-sampling scheme to reduce phylogenetic biases in a MSA}
\index{contact prediction!maximum entropy model!log-likelihood!reweighted-sampling}

In statistical approaches described in this section,
sequences in a MSA are assumed to be
independently and identically distributed samples in sequence space.
However, homologous sequences in a MSA are actually biased 
due to phylogenetic relations between species. 
Such phylogenetic biases are reduced here by taking weighted averages over
the sequences in calculation of single and pairwise amino acid frequencies.
The amino acid frequencies $f_i(a_k)$ at site $i$ and 
pairwise frequencies $f_{ij}(a_k, a_l)$ between sites $i$ and $j$
are calculated as follows.
\begin{eqnarray}
	f_i(a_k) &=& \frac{1}{M_{\script{eff}}} \sum_{\tau} w_{\VEC{\sigma}^{\tau}} \delta_{\sigma^{\tau}_i a_k}
	\hspace*{1em} , \hspace*{1em}
	f_{ij}(a_k,a_l) = \frac{1}{M_{\script{eff}}} \sum_{\tau} w_{\VEC{\sigma}^{\tau}} 
	\delta_{\sigma^{\tau}_i a_k} \delta_{\sigma^{\tau}_j a_l}
	\label{eq: phylogenetic_bias}
	\\
	w_{\VEC{\sigma}} &=& 1 / \, [\, \sum_{\tau} \theta( \sum_{i=1}^L \delta_{\sigma_i \sigma^{\tau}_i} -  s L ) \, ]
	\hspace*{1em} , \hspace*{1em}
	B = M_{\script{eff}} \equiv \sum_{\tau} w_{\VEC{\sigma}^{\tau}}
	\label{eq: weight_for_sequence}
\end{eqnarray}
where $\theta$ is a Heaviside step function, and $s$ is a similarity threshold for sequence identity 
to regard two sequences virtually identical;
$s=0.8$\CITE{MPLBMSZOHW:11} and $s=0.7$\CITE{MCSHPZS:11,WSLZX:17} were employed.
Thus, $B$ in \Eq{\Ref{eq: log-likelihood}} 
is taken here to be equal to $M_{\script{eff}}$ that is the effective number of sequences.
In GaussDCA\CITE{BZFPZWP:14}, the threshold $1 - s$
is defined as being inversely proportional to the average sequence identity
over all pairs of sequences; s = 1 - 0.1216 / \text{average-sequence-identity}.

Except for models in which priors or regularization terms are not taken into account,
these corrected frequencies are employed 
as single-site and two-site marginal probabilities;
\begin{eqnarray}
P_i(a_k) &=& f_i(a_k) \hspace*{2em} P_{ij}(a_k,a_l) = f_{ij}(a_k,a_l)
\end{eqnarray}
In the models with $R=0$, pseudocount based on Bayesian statistics, 
which is a correction scheme for small sample size, is employed 
to estimate $P_i(a_k)$ and 
\LongVersion{
$P_{ij}(a_k,a_l)$; 
see \Eqs{\Ref{eq: pseudo-count_for_Pi} and \Ref{eq: pseudo-count_for_Pij}}.
} 
\ShortVersion{
$P_{ij}(a_k,a_l)$. 
} 

} 

\subsection{Inverse Potts model}
\index{contact prediction!maximum entropy model!inverse Potts model}

\subsubsection{A gauge employed for $h_i(a_k)$ and $J_{ij}(a_k, a_l)$}
\label{section: q-gauge}
\index{contact prediction!maximum entropy model!inverse Potts model!gauge}

Unless specified, a following gauge
is employed; we call it $q$-gauge, here. 
\begin{eqnarray}
        h_i(a_q) &=& J_{ij}(a_k, a_q) = J_{ij}(a_q, a_l) = 0
        \label{eq: q gauge}
\end{eqnarray}
In this gauge,
the amino acid $a_q$ is the reference state for fields and couplings, 
and $P_i(a_q)$, $P_{ij}(a_k,a_q) = P_{ji}(a_q,a_k)$, and $P_{ij}(a_q,a_q)$ 
are regarded as dependent variables.
Common choices for the reference state $a_q$ are
the most common (consensus) state at each site.
Any gauge can be transformed to another by the following transformation.
\begin{eqnarray}
J^{\script{I}}_{ij}(a_k,a_l) &\equiv& J_{ij}(a_k,a_l) - J_{ij}(\cdot,a_l)
			- J_{ij}(a_k, \cdot) +  J_{ij}(\cdot,\cdot)
		\\
h^{\script{I}}_{i}(a_k) &\equiv& h_{i}(a_k) - h_{i}(\cdot) + 
	\sum_{j \neq i} (J_{ij}(a_k, \cdot) - J_{ij}(\cdot, \cdot) )
\end{eqnarray}
where ``$\cdot$'' denotes the reference state, which may be $a_q$ for each site (q-gauge) or
the average over all states (Ising gauge).

\subsubsection{Boltzmann machine}
\label{section: Boltzmann_machine}
\index{contact prediction!maximum entropy model!Boltzmann machine learning}

Fields $h_i(a_k)$ and couplings $J_{ij}(a_k,a_l)$ are
estimated by iterating the following 2-step procedures.
\begin{enumerate}

\item For a given set of $h_i$ and $J_{ij}(a_k, a_l)$, 
marginal probabilities,  
$P^{\script{MC}}(\sigma_i=a_k)$ and $P^{\script{MC}}(\sigma_i=a_k,\sigma_j=a_l)$,
are estimated by a Markov chain Monte Carlo method
(the Metropolis-Hastings algorithm\CITE{MRRT:53}) or by any other method 
(for example, the message passing algorithm\CITE{WWSHH:09}).

\item Then, $h_i$ and $J_{ij}(a_k, a_l)$ are 
updated according to 
the gradient of negative log-posterior-probability per instance,
$\partial S_0/\partial h_i(a_k)$ or $\partial S_0/\partial J_{ij}(a_k,a_l)$,
multiplied by a parameter-specific weight factor\CITE{BLCC:16}, 
$w_i(a_k)$ or $w_{ij}(a_k,a_l)$;
see \Eqs{\Ref{eq: cross-entropy} and \Ref{eq: gradient_of_Z}}.
\begin{eqnarray}
\Delta h_i(a_k) &=& - (P^{\script{MC}}(\sigma_i=a_k) 
	+ \frac{\partial R}{\partial h_i(a_k)} - P_i(a_k)) \cdot w_i(a_k)
	\\
\Delta J_{ij}(a_k,a_l) &=& - 
	(P^{\script{MC}}(\sigma_i=a_k,\sigma_j=a_l) + \frac{\partial R}{\partial J_{ij}(a_k,a_l)} - P_{ij}(a_k,a_l)) \cdot w_{ij}(a_k,a_l)
\end{eqnarray}
where weights are also updated as 
$w_i(a_k) \leftarrow f(w_i(a_k))$ and $w_{ij}(a_k,a_l) \leftarrow f(w_{ij}(a_k,a_l))$
according to the RPROP\CITE{RB:93} algorithm;
the function $f(w)$ is defined as
\begin{eqnarray}
	f(w) &\equiv& \left\{ \begin{array}{ll}
		\max(w \cdot s_{-}, w_{\script{min}}) & \text{ if the gradient changes its sign, }
					\\
		\min(w \cdot s_{+}, w_{\script{max}}) & \text{ otherwise }
				\end{array}
			\right.
\end{eqnarray}
The
$w_{\script{min}}=10^{-3}, w_{\script{max}}=10, s_{-} = 0.5$, and $s_{+} = 1.9 < 1/s_{-} $ were employed\CITE{BLCC:16}.
After updated, $h_i(a_k)$ and $J_{ij}(a_k,a_l)$ may be modified to satisfy a given gauge.

\end{enumerate}
The Boltzmann machine has a merit
that model correlations are calculated.

\LongVersion{

\subsubsection{Message passing algorithm to estimate marginal probabilities 
}
\index{contact prediction!maximum entropy model!message passing algorithm}

To estimate marginal probabilities for given fields and couplings,
the standard belief propagation algorithm for single-site marginals and 
the generalized belief propagation (susceptibility propagation\CITE{MM:08}) 
algorithm 
for two-site marginals were employed in mpDCA\CITE{WWSHH:09}; 
these algorithms are exact for trees but approximate for 
general graphs that is the present case.

\subsubsubsection{Belief propagation algorithm to estimate single-site marginals,
$\{P^{\script{MP}}(\sigma_i=a_k)\}$}

\noindent
In this algorithm for single-site marginals, 
messages (beliefs) $P^{\script{MP}}_{i\rightarrow m}(a_k)$,
which may be understood as the marginal
distribution at site $i$ in the system where site $m$ is removed,
are self-consistently passed between sites.
The self-consistent messages can be obtained by iteratively solving
the following equations with arbitrary initial messages
for the Hamiltonian defined in \Eq{\ref{eq: H_for_Potts}}, i.e., 
given $h^{\script{MP}}$ and $J^{\script{MP}}$.
\begin{eqnarray}
	P^{\script{MP}}_{i\rightarrow m}(a_k) &\propto& \exp(h^{\script{MP}}_i(a_k)) \prod_{j\neq i,m} 
		[ \sum_{l} \exp(J^{\script{MP}}_{ji}(a_l,a_k)) P^{\script{MP}}_{j\rightarrow i}(a_l) ]
	\label{eq: belief}
\end{eqnarray}
with $\sum_{k} P^{\script{MP}}_{i\rightarrow m}(a_k) = 1$.
Random sequential updates seem to be most efficient for solving\CITE{WWSHH:09}.

Having calculated all messages, true single-site marginals
can be estimated by
\begin{eqnarray}
	P^{\script{MP}}(\sigma_i=a_k) &\propto& \exp(h^{\script{MP}}_i(a_k)) \prod_{j \neq i} 
		[ \sum_{l} \exp(J^{\script{MP}}_{ji}(a_l,a_k)) P^{\script{MP}}_{j\rightarrow i}(a_l) ]
	\label{eq: marginal_at_i_in_MP}
\end{eqnarray}

After convergence 
the fields can be determined 
	from \Eq{\ref{eq: marginal_at_i_in_MP}} as
\begin{eqnarray}
	\exp(h^{\script{MP}}_i(a_k)) &\propto&  \frac{ P_i(a_k) }
		{ \prod_{j\neq i} [\, \sum_{l} \exp(J^{\script{MP}}_{ji}(a_l,a_k)) P^{\script{MP}}_{j\rightarrow i}(a_l) } \,]
	\label{eq: fields_in_MP}
\end{eqnarray}
The fields $h^{\script{MP}}_i(a_k)$ must be fixed by a given gauge condition.
\Eq{\ref{eq: fields_in_MP}} guarantees that the single-site marginals agree to given amino acid frequencies 
($P^{\script{MP}}(\sigma_i=a_k) = P_i(a_k)$) 
and therefore the gradient descent update for fields is not needed\CITE{WWSHH:09}.
This characteristic is useful in the inference (gradient descent updates) 
of couplings for Potts models with $q > 3$ that tend to show first-order phase transitions, and
in which small changes in fields and couplings may lead to large changes in marginals\CITE{WWSHH:09}.

\subsubsubsection{Susceptibility propagation algorithm to estimate 
two-site marginals, $\{P^{\script{MP}}(\sigma_i=a_k,\sigma_j=a_l)\}$ }

After all single-site marginals are calculated for a given Hamiltonian,
the covariance matrix, which includes two-site marginals,
can be estimated after iterations to obtain self-consistent messages in
the equation of susceptibility propagation that
is derived from \Eq{\ref{eq: belief}} as follows.
\noindent
\begin{eqnarray}
	M^{\script{MP}}_{i\rightarrow m; j}(a_k,a_l) 
	&\equiv& \frac{\partial P^{\script{MP}}_{i\rightarrow m}(a_k)} {\partial h^{\script{MP}}_j(a_l))}
	=
	P^{\script{MP}}_{i\rightarrow m}(a_k) 
	\, [ \, \frac{\partial \log f_{i\rightarrow m}(a_k)}{\partial h^{\script{MP}}_j(a_l))} 
	- \sum_k P^{\script{MP}}_{i\rightarrow m}(a_k)
	 \frac{\partial \log f_{i\rightarrow m}(a_k)}{\partial h^{\script{MP}}_j(a_l))}
	\, ]
	\label{eq: susceptibility_propagation}
\end{eqnarray}
\begin{eqnarray}
\frac{\partial \log f_{i\rightarrow m}(a_k)}{\partial h^{\script{MP}}_j(a_l))}
	&=& \, [ \, \delta_{ij}\delta_{kl}
		+ \sum_{n \neq i,m}  
		\frac{\sum_{\nu}\exp(J^{\script{MP}}_{ni}(a_{\nu},a_{k}) M^{\script{MP}}_{n\rightarrow i; j}(a_{\nu},a_l)}
		{\sum_{\nu} \exp(J^{\script{MP}}_{ni}(a_{\nu},a_{k}) P^{\script{MP}}_{n\rightarrow i}(a_\nu) }
		\, ]
	\label{eq: susceptibility_propagation_2}
\end{eqnarray}
where the second term of \Eq{\ref{eq: susceptibility_propagation}} 
is a term that results from the partial derivative of the 
normalization of $P^{\script{MP}}_{i\rightarrow m}(a_k)$ in \Eq{\ref{eq: belief}}.
Then, 
similarly
the covariance matrix can be estimated as follows .
\begin{eqnarray}
\frac{\partial P^{\script{MP}}(\sigma_i=a_k)}{\partial h^{\script{MP}}_j(a_l)}
	&=& P^{\script{MP}}(\sigma_i=a_k, \sigma_j=a_l) - P_i(a_k) P_j(a_l)
	\\
	&=& P_{i}(a_k) 
	\, [ \, \frac{\partial \log f_{i}(a_k)}{\partial h^{\script{MP}}_j(a_l))} 
	- \sum_k P_{i}(a_k)
	 \frac{\partial \log f_{i}(a_k)}{\partial h^{\script{MP}}_j(a_l))}
	\, ]
	\\
\frac{\partial \log f_{i}(a_k)}{\partial h^{\script{MP}}_j(a_l))}
	&=& \, [ \, \delta_{ij}\delta_{kl}
		+ \sum_{n \neq i}  
		\frac{\sum_{\nu}\exp(J^{\script{MP}}_{ni}(a_{\nu},a_{k}) M^{\script{MP}}_{n\rightarrow i; j}(a_{\nu},a_l)}
		{\sum_{\nu} \exp(J^{\script{MP}}_{ni}(a_{\nu},a_{k}) P^{\script{MP}}_{n\rightarrow i}(a_\nu) }
		\, ]
\end{eqnarray}
In the equation above, $P^{\script{MP}}(\sigma_i=a_k) = P_i(a_k)$ is employed, because
it is guaranteed by \Eq{\ref{eq: fields_in_MP}}.

It requires very intensive calculations to obtain self-consistent $M^{\script{MP}}_{i\rightarrow m; j}(a_k,a_l)$,
because the number of the messages is $O(L^3q^2)$, and an efficient implementation requires $O(L^4q^2)$ steps of 
calculations for a given set of fields $h^{\script{MP}}_i(a_k)$ and couplings $J^{\script{MP}}_{ij}(a_k,a_l)$\CITE{WWSHH:09}.

} 

\LongVersion{

\subsubsection{Mean field approximation for the inverse Potts model} 
\label{sec: MF}
\index{contact prediction!maximum entropy model!mean field approximation}

In this approximation, 
no prior knowledge ($P_0 = \textrm{constant}$) or
the limit of a infinite number of instances ($B \rightarrow \infty$)
is assumed; the regularization term $R=0$ 
in \Eq{\ref{eq: cross-entropy}}.
$B \rightarrow \infty$ is reasonable for most problems in physics.
For the present case, however, pseudocount
must be employed because of a small number of instances.

\subsubsubsection{$h^{\script{MF}}_i(a_k)$ in the mean field approximation}

The partition function is expanded in terms of coupling $J$ by 
representing it as a function of $\alpha$ as follows\CITE{P:82}.
\begin{eqnarray}
	- F(\alpha) &\equiv& \log Z(\alpha)
	= \log Z^{\script{MF}}(\alpha) + O(\alpha^2 )
\end{eqnarray}
where
\begin{eqnarray}
	Z(\alpha) &\equiv& \sum_{\VEC{\sigma}} \exp \, [ \,
	+ \sum_i h_i(\sigma_i) + \frac{\alpha}{2} \sum_i \sum_{j\neq i}
		J_{ij}(\sigma_i, \sigma_j) \, ]
\end{eqnarray}
Then, ignoring the terms of $\alpha^n$ with $n>1$,  
the partition function in the mean field approximation, which would be
adequate to the case of weak couplings, is obtained.
\begin{eqnarray}
&-& F^{\script{MF}}(1)
	\equiv \log Z^{\script{MF}}(1)
	\nonumber
	\\
	&=& - \sum_i \sum_k P_i(a_k) \log P_i(a_k)
	+ \sum_i \sum_k 
	[ \, 
	h_i(a_k) P_i(a_k)
	+ \frac{1}{2} \sum_{j\neq i} \sum_l J_{ij}(a_k,a_l) P_i(a_k)P_j(a_l)
	\, ]
	\label{eq: Z_of_mean_field_approx}
\end{eqnarray}
$P_i(a_k)$ minimizing $F^{\script{MF}}(P_i(a_k))$ satisfies 
\begin{eqnarray}
	P_i(a_k) &=& 
	\frac{ \exp ( h^{\script{MF}}_i(a_k) + \sum_{j \neq i} \sum_{l \neq q} 
		J^{\script{MF}}_{ij}(a_k,a_l) P_j(a_l) )
		}
		{ 1 + \sum_{k \neq q}
		 \exp( h^{\script{MF}}_i(a_k) + \sum_{j \neq i} \sum_{l \neq q} 
			J^{\script{MF}}_{ij}(a_k,a_l) P_j(a_l) )
		}
	\label{eq: h_of_mean_field_approx}
\end{eqnarray}
This famous mean field equation can be utilized to 
estimate fields $h^{\script{MF}}_i(a_k)$
from marginal probabilities ($P_i(a_k)$) and couplings $J^{\script{MF}}_{ij}(a_k,a_l)$, which are derived in the next paragraph.

\subsubsubsection{$J^{\script{MF}}_{ij}(a_k, a_l)$ in the mean field approximation}

The free energy $F$ for the Potts model satisfies
\begin{eqnarray}
	dF(h, J) &=& - \sum_i \sum_{k \neq q} P_i(a_k) d h_i(a_k) 
		- \sum_i \sum_{k \neq q} \sum_{j>i} \sum_{l \neq q} P_{ij}(a_k, a_l) dJ_{ij}(a_k, a_l)
\end{eqnarray}
We transform variables from $h_i$  to $P_i$ by a Legendre transformation as follows
\CITE{GY:91}.
\begin{eqnarray}
	- G(h, \{P_{ij}\}|\{P_i\},J) 
	&=& \log Z(\{P_i\},J ) - \sum_i \sum_{k \neq q} h_i(a_k) P_i(a_k) 
	\label{eq: Potts_free_energy_of_Pi_Jij}
\end{eqnarray}

The given covariance matrix $C$ can be represented as follows.
\begin{eqnarray}
	C_{i(q-1)+k,n(q-1)+m} &\equiv&
	C_{in}(a_k, a_m) \equiv
	P_{in}(a_k,a_m) - P_i(a_k) P_n(a_m)
	\label{eq: covariance_matrix}
	\\
	&=&
	\frac{\partial^2 -F(h,J)}{\partial h_n(a_m) \partial h_i(a_k)}
	=
	(\frac{\partial P_i(a_k)}{\partial h_n(a_m)})_{\{h_i\},\{J_{ij}\}}
	\\
	(C^{-1})_{i(q-1)+k,n(q-1)+m} 
	&\equiv&
	(C^{-1})_{in}(a_k,a_m)
	= 
	(\frac{\partial^2 G(\{P_i\},J) }{\partial P_i(a_k) \partial P_n(a_m)})	
	=
	(\frac{\partial h_i(a_k)}{\partial P_n(a_m)})_{\{P_i\},\{J_{ij}\}}
	\label{eq: inv_C_in_MF}
\end{eqnarray}
From $h_i(a_k) \simeq h^{\script{MF}}_i(a_k)$, \Eqs{\Ref{eq: h_of_mean_field_approx} and \Ref{eq: inv_C_in_MF}},
the following equation for couplings $J^{\script{MF}}_{ij}(a_k,a_l)$ is derived.
\begin{eqnarray}
	- (C^{-1})_{in}(a_k,a_m) &\simeq&
	J^{\script{MF}}_{in}(a_k,a_m)(1-\delta_{i,n}) - \delta_{in}(\frac{\delta_{km}}{P_n(a_m)} + \frac{1}{P_n(a_q)} ) 
	\label{eq: J_MF}
\end{eqnarray}

} 

\LongVersion{

\subsubsubsection{Pseudocount to make the covariance matrix invertible}
\index{contact prediction!maximum entropy model!mean field approximation!pseudocount}

\noindent
Couplings $J^{\script{MF}}$ in the mean field approximation
is equal to $- C^{-1}$ except diagonal $(q-1) \times (q-1)$ matrices; 
see \Eq{\ref{eq: J_MF}}. 
In the case of protein families, the depth of MSA is very limited, $B < (q-1) L$,
and therefore the observed covariance matrix $C$ is singular and 
must be modified to be regular.
Here, pseudocount based on Bayesian statistics is employed\CITE{WWSHH:09}.
\begin{eqnarray}
	P_i(a_k) &=& (1 - p_c) f_i(a_k) + p_c \frac{1}{q}
	\label{eq: pseudo-count_for_Pi}
	\\
	P_{ij}(a_k,a_l) &=& 
		\left\{ \begin{array}{ll} 
			(1 - p_c) f_{ij}(a_k,a_l) + p_c \frac{1}{q^2} & \text{ for } i \neq j
				\\
			P_i(a_k) \delta_{kl}		& \text{ for } i = j
			\end{array}
			\right.
	\label{eq: pseudo-count_for_Pij}
	\label{eq: pseudo-count}
	\label{eq: pseudocount}
\end{eqnarray}
where $p_c$ is the ratio of pseudocount.
Pseudocount based on Bayesian statistics
depends on the number of samples, $B$, and has a characteristic of 
$p_c \rightarrow 0$ for $B \rightarrow \infty$; 
$p_c = \kappa / (\kappa + B)$.
In the DCA method\CITE{MPLBMSZOHW:11,MCSHPZS:11}, however, $p_c = 0.5$ was employed 
independently of the effective number of sequences, $B$.

} 

\LongVersion{

\subsubsection{Continuous multivariate Gaussian approximation for $P(\VEC{\sigma})$}
\index{contact prediction!maximum entropy model!Gaussian approximation}

The cross-entropy for the Potts is evaluated  
by approximating the distribution of sequences with a multivariate Gaussian distribution,
$P(\VEC{\sigma}) \simeq \mathcal{N}(\{ \delta_{\sigma_i a_k} \} |\{P_i(a_k)\}, \Theta^{-1})$.
In this approximation, $q$-gauge is employed, and 
$P_i(a_q)$, $P_{ij}(a_k,a_q) = P_{ji}(a_q,a_k)$, and $P_{ij}(a_q,a_q)$,
are regarded as dependent variables; see section \ref{section: q-gauge}.
\begin{eqnarray}
S_{\script{Potts}}(\Theta | \{P_i\}, \{P_{ij}\}) &\simeq&
	- \frac{1}{2} \, [ \, 
	- \text{Tr} \, C \Theta + \log \det \Theta - \dim{\Theta} \log 2\pi
	\, ]
	\label{eq: Gaussian_approx}
\end{eqnarray}
where $\Theta$ is a $(q-1)L \times (q-1)L$ precision matrix, which is the inverse covariance matrix,
and $C$ is the given covariance matrix defined by \Eq{\ref{eq: covariance_matrix}}.

In the case of no regularization term, for which
pseudocount (\Eqs{\Ref{eq: pseudo-count_for_Pi} and \Ref{eq: pseudo-count_for_Pij}}) 
must be employed, the estimate of $\Theta$ is given by
\begin{eqnarray}
\Theta &=& \arg \min_{\Theta} S_{\script{Potts}}(\Theta | \{P_i\}, \{P_{ij}\}) = C^{-1}
	\\
S_{\script{Potts}}(\{P_i\}, \{P_{ij}\}) &\simeq& \min_{\Theta} S_{\script{Potts}}(\Theta | \{P_i\}, \{P_{ij}\})
	= - (\log \det C^{-1}) / 2 + \text{constant} 
\end{eqnarray}
Then, couplings and fields are estimated with \Eq{\ref{eq: optimum_h_and_J_from_minimum_S_0}} to be
\begin{eqnarray}
	J^{\script{Gauss}}_{ij}(a_k,a_l) 
	&=&
	- \frac{\partial S_{\script{Potts}}(\{P_i\}, \{P_{ij}\})}{\partial P_{ij}(a_k,a_l) }
	= - (C^{-1})_{ij}(a_k, a_l)
	\label{eq: J_in_Gaussian}
	\\
 h^{\script{Gauss}}_i(a_k) 
	&=& - \sum_{j\neq i} \sum_{l\neq q} J_{ij}(a_k,a_l) P_j(a_l) 
		- \frac{1}{2} \sum_{l\neq q} (C^{-1})_{ii}(a_k,a_l) ( \delta_{kl} - 2 P_i(a_l) ) 
	\label{eq: h_in_Gaussian}
\end{eqnarray}
Thus, the estimate of $J_{ij}(a_k,a_l)$ in the Gaussian approximation 
without regularization terms is 
equivalent to that in the mean field approximation.
The estimates of fields 
are different from those (\Eq{\ref{eq: h_of_mean_field_approx}}) in the mean field approximation,
and give a better generative model\CITE{BLCC:16}.

In the following, 
the $(q-1)L \times (q-1)L$ precision matrix $\Theta$ is estimated by adding regularization terms to
the cross entropy of the inverse Potts model. The estimate of $\Theta$ may be employed instead of
$C^{-1}$ in \Eqs{\Ref{eq: J_in_Gaussian} and \Ref{eq: h_in_Gaussian}} to estimate $h$ and $J$.

\subsubsubsection{With a $\ell_2$ regularization term for precision matrix}

In the following case of a $\ell_2$ regularization term,
the precision matrix $\Theta$
that minimizes the cross-entropy defined 
by \Eqs{\Ref{eq: cross-entropy} and \Ref{eq: Gaussian_approx}} 
can be analytically calculated.
\begin{eqnarray}
\Theta^{\script{L2}} &=& \arg \min_{\Theta} 
	\, [ \, S_{\script{Potts}}(\Theta | \{P_{i}\}, \{P_{ij}\})	
	+ R(\Theta)
	\  ]
	\ , \  
R(\Theta) \equiv  
	\frac{\gamma}{2} \sum_i \sum_{j} \sum_{k \neq q} \sum_{l\neq q} \Theta_{ij}(a_k, a_l)^2
\end{eqnarray}
The optimum $\Theta \,( = \arg \min_{\Theta} S_{0}(\Theta | \{P_i\}, \{P_{ij}\}) )$ is the root of
$C - {\Theta}^{-1} + 2 \gamma \Theta = 0$.
Hence, $\Theta$ and $C$ can be diagonalized by the same orthogonal matrix\CITE{CM:12}.
Let $\theta_{\lambda}$ and $c_{\lambda}$ denote the $\lambda$th eigenvalues 
of $\Theta$ and $C$, respectively.  
$\theta_{\lambda}$ can be explicitly represented as follows.
\begin{eqnarray}
\theta_{\lambda}^{-1} &=& \frac{ 1}{2} ( c_{\lambda} + \sqrt{c_{\lambda}^2 + 8\gamma} )
\end{eqnarray}

\subsubsubsection{With a $\ell_1$ regularization term for precision matrix}

The graphical lasso is employed in PSICOV\CITE{JBCP:12}
to infer the precision matrix $(C^{-1})_{ij}(a_k,a_l)$;
the following function, 
in which a $\ell_1$ regularization term corresponding to an exponential prior 
is employed, 
is minimized.
\begin{eqnarray}
\Theta^{\script{glasso}} &=& \arg \min_{\Theta} 
	\, [ \, S_{\script{Potts}}(\Theta | \{P_{i}\}, \{P_{ij}\})	
	+ R(\Theta)
	\, ]
	\ , \  
R(\Theta) \equiv
	\frac{\gamma}{2} \sum_i \sum_{j} \sum_{k\neq q} \sum_{l \neq q} | \Theta_{ij}(a_k, a_l) |
	\label{eq: graphical_lasso}
\end{eqnarray}
To speed up convergence, a shrinkage method\CITE{SS:05}
for a sample covariance matrix is also employed in PSICOV\CITE{JBCP:12}.
\begin{eqnarray}
	\hat{C} &=& \lambda C + (1 - \lambda) \bar{v} I
	\ , \hspace*{1em}
	\bar{v} = \sum_i \sum_{k \neq q} C_{ii}(a_k, a_k) / (\sum_i \sum_{k \neq q} 1)
\end{eqnarray}
where $0 < \lambda < 1$ is a parameter, and $I$ is the identity matrix.

In COUSCOus\CITE{RMKAAUB:16}, the following shrinkage with
an empirical Bayes estimator is performed 
until the adjusted covariance matrix $\hat{C}$ becomes invertible;
	$\hat{C} = C + \{ (\text{dim}C -1)/(B \text{Tr}C) \} I$.
Then, the adjusted covariance matrix $\hat{C}$ 
is employed 
instead of $C$ in \Eqs{\Ref{eq: Gaussian_approx} and \Ref{eq: graphical_lasso}}.

In PSICOV and COUSCOus, $\Theta$ was supposed to be $qL \times qL$, but 
in the present model
it is $(q-1)L \times (q-1)L$ matrix to consist of
linearly independent rows only even in the limit of $\gamma (\propto 1/B) \rightarrow 0$.

} 

\subsubsection{Gaussian approximation for $P(\VEC{\sigma})$ with a normal-inverse-Wishart prior}
\index{contact prediction!maximum entropy model!Gaussian approximation!normal-inverse-Wishart prior}

\noindent
The normal-inverse-Wishart distribution (NIW) is 
the product of the multivariate normal distribution ($\mathcal{N}$) 
and the inverse-Wishart distribution ($\mathcal{W}^{-1}$), which 
are the conjugate priors for the mean vector and for
the covariance matrix of a multivariate Gaussian distribution, respectively.
The NIW is employed as a prior in GaussDCA\CITE{BZFPZWP:14}, 
in which the sequence distribution $P(\VEC{\sigma})$ is approximated as a Gaussian distribution.
In this approximation, the q-gauge is used, and
$P_i(a_q)$, $P_{ij}(a_k,a_q) = P_{ji}(a_q,a_k)$, and $P_{ij}(a_q,a_q)$
are regarded as dependent variables; see section \ref{section: q-gauge};
in GaussDCA, deletion is excluded from independent variables.

The posterior distribution for the NIW is also a NIW.
Thus, the cross entropy $S_0$ can be represented as
\begin{eqnarray}
\lefteqn{
S_0(\VEC{\mu}, \Sigma| \{P_i\}, \{P_{ij}\}) 
=
		\frac{-1}{B}\log 
		[\, \prod_{\tau=1}^B \mathcal{N}(\{ \delta_{\sigma^{\tau}_i a_k} \} |\VEC{\mu}, \Sigma)
		 \mathcal{N}(\VEC{\mu}|\VEC{\mu}^{0}, \Sigma / \kappa )
		\mathcal{W}^{-1}( \Sigma | \Lambda, \nu ) \, ]
}
		\\
&=&
		\frac{-1}{B}\log 
		[\, \mathcal{N}(\VEC{\mu}|\VEC{\mu}^{B}, \Sigma / \kappa^{B})
		\mathcal{W}^{-1}( \Sigma | \Lambda^{B}, \nu^{B} ) 
		\\
		& &
		(\det (2\pi \Sigma))^{- B/2} (\frac{\kappa}{\kappa^B})^{\dim{\Sigma}/2}
		\frac{(\det (\Lambda/2) )^{\nu/2}}{(\det (\Lambda^B/2) )^{\nu^B/2}}
		\frac{\Gamma_{\dim\Sigma}(\nu^B/2)}{\Gamma_{\dim\Sigma}(\nu/2)}
		(\det \Sigma)^{-(\nu - \nu^B)2}
		\, ]
\end{eqnarray}
where
$\Gamma_{\dim{\Sigma}}(\nu/2)$ is the multivariate $\Gamma$ function, 
$\VEC{\mu}$ is the mean vector, and $\dim{\Sigma}$
is the dimension of covariance matrix $\Sigma$, $\dim{\Sigma} = (q-1) L$ excluding deletion 
in GaussDCA. 
The normal and NIW distributions are defined as follows.
\begin{eqnarray}
\mathcal{N}(\VEC{\mu}|\VEC{\mu}^0, \Sigma)
	&\equiv& (\det (2\pi \Sigma))^{-1/2}
	\exp(- \frac{(\VEC{\mu} - \VEC{\mu}^0)^T \Sigma^{-1}(\VEC{\mu} - \VEC{\mu}^0 )}{2} )
	\label{eq: def_normal_distribution}
	\\
\mathcal{W}^{-1}( \Sigma | \Lambda, \nu ) &\equiv&
	\frac{ (\det (\Lambda/2))^{\nu/2}}
		{\Gamma_{\dim{\Sigma}}(\nu/2) }
	 (\det \Sigma)^{-(\nu + \dim{\Sigma} + 1)/2} \exp(- \frac{1}{2} \text{Tr} \Lambda \Sigma^{-1} )
	\label{eq: def_inverse_Wishart_distribution}
\end{eqnarray}
Parameters
$\VEC{\mu}^{B}$, $\kappa^{B}$, $\nu^{B}$, and $\Lambda^{B}$
satisfy
\begin{eqnarray}
	\mu^{B}_i(a_k) &=& (\kappa \mu^0_i(a_k) + B P_i(a_k)) / (\kappa + B)
	\  , \ 
	\kappa^{B} = \kappa + B
	\  , \ 
	\nu^{B} = \nu + B
	\\
	\Lambda^{B}_{ij}(a_k,a_l) &=& 
	\Lambda_{ij}(a_k,a_l) + B C_{ij}(a_k,a_l)  + \frac{\kappa B}{\kappa + B} [ (P_i(a_k) - \mu^0_i(a_k)) (P_j(a_l) - \mu^0_j(a_l)) ]
\end{eqnarray}
where the $\Lambda$ and $\nu$ are the scale matrix and the degree of freedom,
respectively, shaping the inverse-Wishart distribution,
and 
\LongVersion{
$C$ is the given covariance matrix defined by \Eq{\ref{eq: covariance_matrix}}.
} 
\ShortVersion{
$C$ is the given covariance matrix; $C_{ij}(a_k,a_l) \equiv P_{ij}(a_k,a_l) - P_i(a_k) P_j(a_l)$.
} 
The mean values of $\VEC{\mu}$ and $\Sigma$ under NW posterior are 
$\VEC{\mu}^{B}$ and $\Lambda^{B} / (\nu^{B} - \dim{\Sigma} - 1)$, 
and their mode values are $\VEC{\mu}^{B}$ and $\Lambda^{B} / (\nu^{B} + \dim{\Sigma} + 1)$,
which minimize the cross entropy or maximize the posterior probability.
The covariance matrix $\Sigma$ can be estimated to be the exactly same value
by adjusting the value of $\nu$,
whichever the mean posterior or the maximum posterior is employed for the estimation of $\Sigma$.
In GaussDCA, the mean posterior estimate was employed 
but here the maximum posterior estimate is employed according to the present formalism.
\begin{eqnarray}
(\VEC{\mu}, \Sigma) &=& \arg \min_{(\VEC{\mu}, \Sigma)} S_0(\VEC{\mu}, \Sigma | \{P_i\}, \{P_{ij}\})
	= 
	(\VEC{\mu}^{B}, \Lambda^{B} / (\nu^{B} + \dim{\Sigma} + 1) )
	\label{eq: modes_for_NIW}
\end{eqnarray}

According to GaussDCA, 
$\nu$ is chosen in such a way that
$\Sigma_{ij}(a_k,a_l)$ is nearly equal to the covariance matrix
corrected 
\LongVersion{
by pseudocount in \Eq{\ref{eq: pseudo-count_for_Pij}};
} 
\ShortVersion{
by pseudocount;
} 
$\nu = \kappa + \dim{\Sigma} + 1$ for the mean posterior estimate in GaussDCA,
but $\nu = \kappa - \dim{\Sigma} - 1$ for the maximum posterior estimate here.

From \Eq{\ref{eq: optimum_h_and_J_from_minimum_S_0}}, the estimates of couplings and fields are calculated. 
\begin{eqnarray}
	J^{\script{NIW}}_{ij}(a_k,a_l) 
	&=& - \frac{\partial S_0(\{P_i\}, \{P_{ij}\})} { \partial P_{ij}(a_k,a_l) }
	= - \frac{(\kappa + B + 1)}{ \kappa + B } (\Sigma^{-1})_{ij}(a_k,a_l)
	\label{eq: J_NIW}
\end{eqnarray}
Because the number of instances is far greater than 1 ($B \gg 1$),
these estimates of couplings are practically equal to the estimates ($J^{\script{MF}} = - \Sigma^{-1}$)
in the mean field approximation, which was employed in GaussDCA\CITE{BZFPZWP:14}.

\begin{eqnarray}
	h^{\script{NIW}}_{i}(a_k) 
	&=& - \sum_{j \neq i}\sum_l J^{\script{NIW}}_{ij}(a_k,a_l) P_j(a_l)
	- \frac{(\kappa + B + 1)}{\kappa + B}
		\sum_j \sum_{l\neq q}
		(\Sigma^{-1})_{ij}(a_k,a_l)
	\nonumber
	\\
	& &
		\, [ \,
		\delta_{ij}\frac{\delta_{kl} - 2 P_i(a_l)}{ 2}
		+ \frac{\kappa B}{\kappa + B } (P_j(a_l) - \mu^0_j(a_l))
		 \, ]
	\label{eq: h_NIW}
\end{eqnarray}
The $(h^{\script{NIW}}_i(a_k) - h^{\script{NIW}}_i(a_q))$ does not
converge to $\log P_i(a_k)/P_i(a_q)$ as $J^{\script{NIW}} \rightarrow 0$ 
but $h^{\script{MF}}_i(a_k) - h^{\script{MF}}_i(a_q)$ does;
in other words, the mean field approximation gives a better $h$ 
for the limiting case of no couplings
than the present approximation.
Barton et al.\CITE{BLCC:16} reported that the Gaussian approximation
generally gave a better generative model than the mean field approximation.

In GaussDCA\CITE{BZFPZWP:14}, $\VEC{\mu}^0$ and $\Lambda / \kappa$ were chosen to be as uninformative as
possible, i.e., mean and covariance for a uniform distribution.
\begin{eqnarray}
	\mu^0_i(a_k) &=& 1 / q
	\hspace*{1em} , \hspace*{1em}
	\frac{\Lambda_{ij}(a_k, a_l)}{\kappa} 
	= \frac{\delta_{ij}}{q}(\delta_{kl} - \frac{1}{q})
\end{eqnarray}

\subsubsection{Pseudo-likelihood approximation}
\index{contact prediction!maximum entropy model!pseudo-likelihood approximation}

\subsubsubsection{Symmetric pseudo-likelihood maximization}

\noindent
The probability of an instance $\VEC{\sigma}^{\tau}$ is approximated
as follows
by the product of conditional probabilities of observing $\sigma^{\tau}_i$ under the given 
observations $\sigma^{\tau}_{j\neq i}$ of all other sites.
\begin{eqnarray}
	P(\VEC{\sigma}^{\tau}) 
	&\approx& 
	\prod_i P( \sigma_i = \sigma_{i}^{\tau} \, | \, \{\sigma_{j\neq i}=\sigma_{j}^{\tau} \} )
\end{eqnarray}
Then, cross entropy is approximated as 
\begin{eqnarray}
	S_{0}(h, J | \{P_i\},\{P_{ij}\})
	&\approx& 
	S^{\script{PLM}}_{0}(h, J | \{P_i\},\{P_{ij}\})
	\equiv 
	\sum_i S_{0,i}(h, J | \{P_i\},\{P_{ij}\}) 
	\\
S_{0,i}(h, J | \{P_i\},\{P_{ij}\})
	&\equiv&
	\frac{-1}{B} \sum_{\tau} 
\ell_i(\sigma_i = \sigma_{i}^{\tau} \, | \, \{\sigma_{j\neq i}=\sigma_{j}^{\tau} \}, h, J)
	+ R_i(h, J)
\end{eqnarray}
where conditional log-likelihoods and $\ell_2$ norm regularization terms employed
in \CITE{ELLWA:13} are
\begin{eqnarray}
\ell_i(\sigma_{i} = \sigma_{i}^{\tau} \, | \, \{\sigma_{j\neq i}=\sigma_{j}^{\tau} \}, h, J)
	&=&
	\log \, [ \,  
	\frac{\exp(h_i(\sigma_{i}^{\tau}) + \sum_{j\neq i} J_{ij}(\sigma_{i}^{\tau}, \sigma_{j}^{\tau})) }
	{\sum_k \exp( h_i(a_k) + \sum_{j\neq i} J_{ij}(a_k, \sigma_{j}^{\tau}) ) }
	\, ]
	\\
R_i(h, J) &\equiv&
	\gamma_h \sum_k h_i(a_k)^2 + \frac{\gamma_J}{2} \sum_k \sum_{j \neq i} \sum_l J_{ij}(a_k, a_l)^2
\end{eqnarray}
The optimum fields and couplings in this approximation are estimated by minimizing 
the pseudo-cross-entropy, $S^{\script{PLM}}_0$.
\begin{eqnarray}
(h^{\script{PLM}}, J^{\script{PLM}}) 
	&=& \arg \min_{h,J} S^{\script{PLM}}_{0}(h, J | \{P_i\},\{P_{ij}\})
	\label{eq: symmetric pseudo-likelihood maximization}
\end{eqnarray}
\Eq{\ref{eq: symmetric pseudo-likelihood maximization}} is not invariant 
under gauge transformation; the $\ell_2$ norm regularization terms 
in \Eq{\ref{eq: symmetric pseudo-likelihood maximization}} favors 
only a specific gauge that corresponds to
$\gamma_J \sum_l J_{ij}(a_k, a_l) = \gamma_h h_i(a_k)$,
$\gamma_J \sum_k J_{ij}(a_k, a_l) = \gamma_h h_j(a_l)$,
and $\sum_k h_i(a_k) = 0$ for all $i$, $j (> i)$, $k$ and $l$ \CITE{ELLWA:13}.
$\gamma_J = \gamma_h = 0.01$ that is relatively a large value independent of 
$B$ was employed in \CITE{ELLWA:13}. 
$\gamma_h = 0.01$ but $\gamma_J = q (L -1) \gamma_h$ were employed in \CITE{HIPSSSM:17}, in which
gapped sites in each sequence were excluded in the calculation of the Hamiltonian $H(\VEC{\sigma})$,
and therefore $q=20$.

GREMLIN\CITE{KOB:13} employs Gaussian prior probabilities that depend on site pairs.
\begin{eqnarray}
	R_i(h, J) &\equiv&
	\gamma_h \sum_k h_i(a_k)^2 + \sum_k \sum_{j \neq i} \frac{\gamma_{ij}}{2} \sum_l J_{ij}(a_k, a_l)^2
	\\
	\gamma_{ij} &\equiv&
	\gamma_c (1 - \gamma_p \log(P^0_{ij}) )
\end{eqnarray}
where $P^0_{ij}$ is the prior probability of site pair $(i,j)$ being in contact.

\subsubsubsection{Asymmetric pseudo-likelihood maximization}

\noindent
To speed up the minimization of $S_0$, 
a further approximation, in which
$S_{0,i}$ is separately minimized, 
is employed\CITE{EHA:14}, and fields and couplings are estimated as follows.
\begin{eqnarray}
	J^{\script{PLM}}_{ij}(a_k, a_l) 
	&\simeq& \frac{1}{2} ( J^{*}_{ij}(a_k, a_l) + J^{*}_{ji}(a_l, a_k))
	\label{eq: asymmetric pseudo-likelihood maximization}
	\\
(h^{\script{PLM}}_i, J^{*}_{i}) 
	&=& \arg \min_{h_i, J_{i}} 
	S_{0,i}(h, J | \{P_i\},\{P_{ij}\})
\end{eqnarray}
It is appropriate to transform $h$ and $J$ estimated above into
a some specific gauge such as the Ising gauge.

\subsubsection{ACE (adaptive cluster expansion) of cross-entropy for sparse Markov random field}
\index{contact prediction!maximum entropy model!adaptive cluster expansion}

\noindent
The cross entropy $S_0(\{h_i, J_{ij}\}|\{P_i\}, \{P_{ij}\}, i,j \in \Gamma)$
of a cluster of sites $\Gamma$,
which is defined as the negative
log-likelihood per instance in \Eq{\ref{eq: minimum_cross_entropy}},
is approximately minimized by taking account of
sets $L_k(t)$ of only significant clusters consisting of $k$ sites, 
the incremental entropy (cluster cross entropy) $\Delta S_{\Gamma}$ of which is
significant ($|\Delta S_{\Gamma}| > t$) \CITE{CM:11,CM:12,BLCC:16}.
\begin{eqnarray}
S_{0}(\{P_i, P_{ij}| i, j \in \Gamma\})
	&\simeq&
	\sum_{l=1}^{|\Gamma|} 
	\sum_{\Gamma^\prime \in L_{l}(t), \Gamma^\prime \subset \Gamma } 
	\Delta S_{0}(\{P_i, P_{ij}| i, j \in \Gamma^\prime\})
	\label{eq: ACE}
	\\
\Delta S_{0}(\{P_i, P_{ij}| i, j \in \Gamma\}) 
	&\equiv&
	S_{0}(\{P_i, P_{ij}| i, j \in \Gamma\}) -
	\sum_{\Gamma^\prime \subset \Gamma } 
	\Delta S_{0}(\{P_i, P_{ij}| i, j \in \Gamma^\prime\})
	\\
	&=&
	\sum_{\Gamma^\prime \subseteq \Gamma } 
	(-1)^{|\Gamma| - |\Gamma^\prime|}
	S_{0}(\{P_i, P_{ij}| i, j \in \Gamma^\prime\})
	\label{eq: CE}
\end{eqnarray}
$L_{k+1}(t)$ is constructed from $L_{k}(t)$ by
adding a cluster $\Gamma$ consisting of $(k+1)$ sites in a lax case 
provided that any pair of size $k$ clusters 
$\Gamma^1, \Gamma^2 \in L_k(t)$ and $\Gamma^1 \bigcup \Gamma^2 = \Gamma$ 
or in a strict case 
if $\Gamma^\prime \in L_k(t)$ for $\forall \Gamma^\prime$ 
such that $\Gamma^\prime \subset \Gamma$ and $|\Gamma^\prime| = k$.
Thus, \Eq{\ref{eq: ACE}} yields sparse solutions.
The cross entropies $S_{0}(\{P_i, P_{ij}| i, j \in \Gamma^\prime\})$
for the small size of clusters are estimated by minimizing 
$S_{0}( \{h_i, J_{ij}\} | \{P_i, P_{ij}\}, i, j \in \Gamma^\prime)$
with respect to fields and couplings.
Starting from a large value of the threshold $t$ (typically $t=1$),
the cross-entropy $S_0(\{P_i, P_{ij}\}|i,j \in \{1,\ldots,N\})$ is calculated
by gradually decreasing $t$ until its value converges. 
Convergence of the algorithm may also be more difficult for alignments of long
proteins or those with very strong interactions. In such cases, strong
regularization may be employed.

The following regularization terms of $\ell_2$ norm are employed in ACE\CITE{BLCC:16}, 
and so \Eq{\ref{eq: ACE}} is not invariant under gauge transformation.
\begin{eqnarray}
- \frac{1}{B}\log P_0(h,J| i, j \in \Gamma) 
	&=& \gamma_h \sum_{i\in \Gamma} \sum_{k} h_i(a_k)^2 
	+ \gamma_J \sum_{i\in \Gamma} \sum_{k} 
	\sum_{j>i, j\in \Gamma} \sum_{l} J_{ij}(a_k, a_l)^2
\end{eqnarray}
$\gamma_h = \gamma_J \propto 1/B$ was employed\CITE{BLCC:16}.

The compression of the number of Potts states, $q_i \leq q$, 
at each site can be taken into account. All infrequently observed states 
or states that insignificantly contribute to site entropy
can be treated as the same state, and
a complete model can be recovered\CITE{BLCC:16}
by setting $h_i(a_k) = h_i(a_{k^{\prime}}) + 
\log (P_i(a_k)/P^{\prime}_i(a_{k^{\prime}}))$,
and $J_{ij}(a_k, a_l) = J^{\prime}_{ij}(a_{k^{\prime}},a_{l^{\prime}})$, where
``$\prime$'' denotes a corresponding aggregated state and a potential.

Starting from the output set of the fields $h_i(a_k)$ and 
couplings $J_{ij}(a_k,a_l)$ obtained 
from the cluster expansion of the cross-entropy,
a Boltzmann machine is trained 
with $P_i(a_k)$ and $P_{ij}(a_k)$ by the RPROP algorithm\CITE{RB:93} to
refine the parameter values of $h_i$ and $J_{ij}(a_k, a_l)$ \CITE{BLCC:16};
see section \ref{section: Boltzmann_machine}.
This post-processing is also useful
because model correlations are calculated.

An appropriate value of the regularization parameter for trypsin inhibitor
were much larger ($\gamma=1$) for contact prediction 
than those ($\gamma=2/B=10^{-3}$) for recovering true fields and couplings\CITE{BLCC:16}, 
probably because the task of contact prediction
requires the relative ranking of interactions rather than
their actual values.

\subsubsection{Scoring methods for contact prediction}
\index{contact prediction!maximum entropy model!scoring methods}

\LongVersion{
\subsubsubsection{Conditional Mutual Information, $\mathcal{S}^{\script{CMI}}_{ij}$}
\label{sec: CMI}

\noindent

Conditional mutual information for each site pair defined as
follows is calculated by the Monte Carlo 
importance sampling,
and employed as a score to predict residue-residue contacts\CITE{LGJ:02,LGJ:12}.
\begin{eqnarray}
\mathcal{S}^{\script{CMI}}_{ij} &\equiv& 
	\sum_{\sigma_{m\neq i,j}} 
	\sum_k \sum_l 
	P(\sigma_i=a_k, \sigma_j=a_l , \{\sigma_{m \neq i,j}\})
	\log \frac{ P(\sigma_i=a_k, \sigma_j=a_l | \{\sigma_{m \neq i,j}\})} 
		{ P(\sigma_i=a_k | \{\sigma_{m \neq i,j}\}) 
		P(\sigma_j=a_l | \{\sigma_{m \neq i,j}\}) }
	\label{eq: conditional_mutual_information}
\end{eqnarray}
} 

\LongVersion{

\subsubsubsection{Direct information, $\mathcal{S}^{\script{DI}}_{ij}$}
\label{sec: DI}

\noindent
The message passing DCA (mpDCA)\CITE{WWSHH:09} and
the mean field DCA (mfDCA)\CITE{MPLBMSZOHW:11,MCSHPZS:11}
methods
employ direct information
$\mathcal{S}^{\script{DI}}_{ij}$ 
defined as follows for scoring; Sander's group renamed this score
evolutionary coupling (EC)\CITE{HCSRSM:12}.
\begin{eqnarray}
	\mathcal{S}^{\script{EC}}_{ij} 
	&\equiv&
	\mathcal{S}^{\script{DI}}_{ij} 
	\equiv \sum_k \sum_l 
		P^{\script{DI}}_{ij}(a_k,a_l) 
		\log \frac{P^{\script{DI}}_{ij}(a_k,a_l)}{P_i(a_k) P_j(a_l)}
	\label{eq: DI_score}
	\\
	P^{\script{DI}}_{ij}(a_k,a_l) &\propto& 
		\exp(h^{\prime}_{ij}(a_k) + h^{\prime}_{ji}(a_l) + J_{ij}(a_k,a_l))
\end{eqnarray}
where $h^{\prime}_{ij}(a_k)$ and $h^{\prime}_{ji}(a_l)$ are determined
to satisfy the following equations
in the two-site model. 
\begin{eqnarray}
	\sum_l P^{\script{DI}}_{ij}(a_k,a_l) &=&
		\sum_l P^{\script{DI}}_{ji}(a_l,a_k) = P_i(a_k) 
	\hspace*{1em} \text{ for } \hspace*{1em} \forall i \text{ and } \forall k
\end{eqnarray}
A nice characteristic of $S^{\script{DI}}_{ij}$ is its invariance 
with respect to the gauge freedom of the Potts model,
but pseudocount is required to regularize frequencies $P_i(a_k)$ \CITE{ELLWA:13}.

} 

\subsubsubsection{Corrected Frobenius norm ($L_{22}$ matrix norm), $\mathcal{S}^{\script{CFN}}_{ij}$ }

\noindent
For scoring, plmDCA\CITE{ELLWA:13,EHA:14} employs 
the corrected Frobenius norm of $J^{\script{I}}_{ij}$ transformed in the Ising gauge, 
in which $J^{\script{I}}_{ij}$ does not contain anything that could have been explained 
by fields $h_i$ and $h_j$;
$
J^{\script{I}}_{ij}(a_k,a_l) \equiv J_{ij}(a_k,a_l) - J_{ij}(\cdot,a_l)
			- J_{ij}(a_k, \cdot) +  J_{ij}(\cdot,\cdot)
$
where
$J_{ij}(\cdot,a_l) =
		J_{ji}(a_l,\cdot) 
	\equiv \sum_{k=1}^q J_{ij}(a_k,a_l) / q
$.
\begin{eqnarray}
	\mathcal{S}^{\script{CFN}}_{ij} &\equiv&
	  \mathcal{S}^{\script{FN}}_{ij} 
	  - \mathcal{S}^{\script{FN}}_{\cdot j} \mathcal{S}^{\script{FN}}_{i \cdot} / \mathcal{S}^{\script{FN}}_{\cdot \cdot}
	\label{eq: CFN_score}
	\hspace*{1em} , \hspace*{1em}
	\mathcal{S}^{\script{FN}}_{ij} \equiv
		\sqrt{\sum_{k\neq \script{gap}} \sum_{l\neq \script{gap}}  J^{\script{I}}_{ij}(a_k,a_l)^2}
	\label{eq: FN_score}
\end{eqnarray}
where "$\cdot$" denotes average over the indicated variable.  
This CFN score with the gap state excluded in \Eq{\ref{eq: FN_score}} 
performs better\CITE{EHA:14,BZFPZWP:14} than
both scores of FN and 
\LongVersion{
DI/EC defined in \Eq{\ref{eq: DI_score}}.
} 
\ShortVersion{
DI/EC\CITE{WWSHH:09,MPLBMSZOHW:11,MCSHPZS:11,HCSRSM:12}.
} 

\LongVersion{

\subsubsubsection{Corrected $L_{11}$ matrix norm}
\label{sec: L11_norm}

\noindent
In PSICOV\CITE{JBCP:12} and COUSCOus\CITE{RMKAAUB:16}, 
the following corrected $L_{11}$ matrix norm  
is employed.
\begin{eqnarray}
	\mathcal{S}^{\script{PSICOV}}_{ij} &\equiv&
	  \mathcal{S}^{\script{L11}}_{ij} 
	  - \mathcal{S}^{\script{L11}}_{\cdot j} \mathcal{S}^{\script{L11}}_{i \cdot} / \mathcal{S}^{\script{L11}}_{\cdot \cdot}
	\hspace*{1em} , \hspace*{1em}
	\mathcal{S}^{\script{L11}}_{ij} \equiv
	\sum_{k} \sum_{l} | \Theta_{ij}(a_k,a_l) | 
\end{eqnarray}
This type of correction was first employed in \CITE{DWG:08} in order to
reduce entropic and phylogenetic biases.
In PSICOV, this corrected score is converted into an estimated positive
predictive value (PPV) by fitting a logistic function to the observed distribution of
scores\CITE{JBCP:12}.

} 
\LongVersion{

\subsection{Partial correlation of amino acid covariations between sites}
\index{contact prediction!partial correlation of amino acid covariations}

Direct information was defined\CITE{TS:11} in
the similar form to partial correlation coefficient as
\begin{eqnarray}
	\text{DI}^{\script{TS}}_{ij} &\equiv& (\text{MIr}^{-1})_{ij} \, / \, \sqrt{(\text{MIr}^{-1})_{ii} (\text{MIr}^{-1})_{jj}}
	\hspace*{1em} , \hspace*{1em} 
	\text{MIr}_{ij} \equiv \text{MI}_{ij} / S_{ij}
\end{eqnarray}
\begin{eqnarray}
	\text{MI}_{ij} &=& S_i + S_j - S_{ij}
	\: , \:
	S_i \equiv - \sum_{k=1}^{q} P_i(a_k) \log P_i(a_k)
	\: , \:
	S_{ij} \equiv -  \sum_{k=1}^{q} \sum_{l=1}^{q} P_{ij}(a_k) \log P_{ij}(a_k,a_l)
\end{eqnarray}
where $\text{MI}$ is a mutual information, and 
$\text{MIr}$ is the normalized $\text{MI}$.
The pseudocount method of \Eq{\ref{eq: pseudo-count}} 
is employed to make $\text{MIr}$ invertible;
the ratio of pseudocount $p_c=1/(B+1)$ was employed\CITE{TS:11}.

} 

\LongVersion{

\subsection{Partial correlation of amino acid cosubstitutions between sites in protein evolution}
\index{contact prediction!partial correlation of amino acid cosubstitutions}

\subsubsection{Mean of characteristic changes accompanied by substitutions at each site in
each branch of a phylogenetic tree in a maximum likelihood model}

Amino acid substitutions are approximated to occur independently at each site.
Then, if substitutions are assumed to be in the equilibrium state of a time-reversible Markov process,
a likelihood $P(\mathcal{A}_i | T, \Theta, \theta_{\alpha})$ of site $i$
in a multiple sequence alignment (MSA) $\mathcal{A}$ in a phylogenetic tree $T$ 
under a evolutionary model $\Theta$ with a parameter $\theta_{\alpha}$ 
for the variation of selective constraints\CITE{M:11a,M:11b,M:13b} can be calculated by taking any node as a root node.
Let us assume here that the root node is a left node ($v_{bL}$) of a branch $b$.
\begin{eqnarray}
        P(\mathcal{A}_i | T, \Theta, \theta_{\alpha}) &=&
                \sum_{\kappa} \sum_{\lambda}
                P(\mathcal{A}_i, v_{bL} = \kappa, v_{bR} = \lambda | T, \Theta, \theta_{\alpha})
\end{eqnarray}
where 
depending on the evolutionary model
$\kappa$ and $\lambda$ 
correspond to the type of codon or amino acid.
The likelihood for $v_{bL} = \kappa$ and $v_{bR} = \lambda$ at site $i$,
$P(\mathcal{A}_i, v_{bL} = \kappa, v_{bR} = \lambda | T, \Theta, \theta_{\alpha})$,
is calculated as follows \CITE{F:81}.
\begin{eqnarray}
\lefteqn{
        P(\mathcal{A}_i, v_{bL} = \kappa, v_{bR} = \lambda | T, \Theta, \theta_{\alpha}) \equiv
}
        \nonumber \\
        & &
        P_{bL}(\mathcal{A}_i | v_{bL} = \kappa, T, \Theta, \theta_{\alpha})
                f_{\kappa} P(\lambda | \kappa, t_b, \Theta, \theta_{\alpha})
                P_{bR}(\mathcal{A}_i | v_{bR} = \lambda, T, \Theta, \theta_{\alpha})
\end{eqnarray}
where 
$P_{bL}(\mathcal{A}_i | v_{bL} = \kappa, T, \Theta, \theta_{\alpha})$ is
a conditional likelihood of the left subtree with $v_{bL} = \kappa$, and
$f_{\kappa}$ is the equilibrium frequency of $\kappa$,
$P(\lambda | \kappa, t_b, \Theta, \theta_{\alpha})$ is a substitution probability from $\kappa$ to $\lambda$
at the branch $b$ whose length is equal to $t_b$.
In the maximum likelihood (ML) method for phylogenetic trees, the tree $T$ and
parameters $\Theta$ are estimated by maximizing the likelihood; 
$(\hat{T}, \hat{\Theta}) = \arg \max_{T, \Theta} P(\mathcal{A} | T, \Theta)$, where
$P(\mathcal{A} | T, \Theta) = \sum_{\alpha} \prod_i P(\mathcal{A}_i | T, \Theta, \theta_{\alpha}) P(\theta_{\alpha})$ and 
$P(\theta_{\alpha})$ is a prior probability.

The mean $\Delta_{ib}$ of any quantity
$\Delta_{\kappa\lambda}$
accompanied by substitutions from $\kappa$ to  $\lambda$
at each site $i$ in each branch $b$ of a phylogenetic tree
can be calculated as follows; $\Delta_{\kappa\lambda}$
corresponds to characteristic changes for coevolution such as volume and charge changes due to
amino acid substitutions.
\begin{eqnarray}
        \Delta_{ib}(\mathcal{A}_i, \hat{T}, \hat{\Theta}, \theta_{\alpha}) &\equiv& \sum_{\kappa, \lambda}
                \frac{
                \Delta_{\kappa \lambda}
                P(\mathcal{A}_i, v_{bL} = \kappa, v_{bR} = \lambda | \hat{T}, \hat{\Theta}, \theta_{\alpha})
                }
                {
                P(\mathcal{A}_i | \hat{T}, \hat{\Theta}, \theta_{\alpha})
                }
                \label{eq: mean_of_delta}
                \\
        \Delta_{ib}(\mathcal{A}_i, \hat{T}, \hat{\Theta})
                &=& \sum_{\theta_{\alpha}} \Delta_{ib}(\mathcal{A}_i, \hat{T}, \hat{\Theta}, \theta_{\alpha})
                        P(\theta_{\alpha} | \mathcal{A}_i, \hat{T}, \hat{\Theta})
                \label{eq: posterior_mean_of_delta}
\end{eqnarray}
where $P(\theta_{\alpha} | \mathcal{A}_i, \hat{T}, \hat{\Theta})$ 
is a posterior probability;
$P(\theta_{\alpha} | \mathcal{A}_i, \hat{T}, \hat{\Theta}) = P(\mathcal{A}_i | \hat{T}, \hat{\Theta}, \theta_{\alpha}) P(\theta_{\alpha}) / P(\mathcal{A}_i | \hat{T}, \hat{\Theta})$.
A Bayesian method for mapping mutations on a phylogenetic tree was first
discussed by Nielsen\CITE{N:02}, and
the present formulation of
\Eqs{\ref{eq: mean_of_delta} and \ref{eq: posterior_mean_of_delta}}
was introduced as a substitution vector along branches at site $i$
by Dutheil et al. \CITE{DPJG:05}
for detecting coevolving positions in a molecule.
The method named substitution mapping for mapping evolutionary
trajectories of discrete traits on phylogenies
was further extended \CITE{MS:08a,MS:08b,TH:11},
and was shown to provide extremely robust statistics\CITE{OMS:09,RFGDBDR:12}.

\subsubsection{Partial correlation coefficients of feature vectors between sites}

If $\Delta_{\kappa\lambda}$ is defined to be equal to $1$ for $\kappa \neq \lambda$
and $0$ for $\kappa = \lambda$,
$\Delta_{ib}(\mathcal{A}_i, \hat{T}, \hat{\Theta})$
will represent the expected value of substitution probability
at site $i$ in branch $b$.
Let us define a vector $\VEC{\Delta}_i$ as follows, and
consider the correlation of the two vectors,
$\VEC{\Delta}_i$ and $\VEC{\Delta}_j$.
\begin{eqnarray}
\VEC{\Delta}_{i} &\equiv& ( \ldots \; , \; \Delta_{ib}(\mathcal{A}_i, \hat{T}, 
	\hat{\Theta}) - 
	\sum_b \Delta_{ib}(\mathcal{A}_i, \hat{T}, \hat{\Theta}) / \sum_b 1 \; , \; \ldots )^{T}
	\label{eq: substitution_vector}
\end{eqnarray}
where $T$ denotes transpose.

The correlation between sites $i$ and $j$ may be
an indirect correlation resulting from
correlations between sites $i$ and $k$ and between sites $k$ and $j$.
To reduce such indirect correlations, partial correlation coefficients are
employed here.
The partial correlation coefficient is the correlation coefficient between
residual vectors
($\Pi_{\bot\{\Delta_{k\neq i, j}\}} \VEC{\Delta}_i$ and $\Pi_{\bot\{\Delta_{k\neq i, j}\}} \VEC{\Delta}_j$)
of given two vectors that are perpendicular to a subspace
consisting of other vectors except those two vectors ($\VEC{\Delta}_i$ and $\VEC{\Delta}_j$)
and therefore cannot be accounted for
by a linear multiple regression on other vectors;
$\Pi_{\bot\{\Delta_{k\neq i, j}\}}$ is a projection operator to a space perpendicular to the subspace.
If the correlation matrix $C$ is invertible, then the partial correlation coefficients $\mathcal{C}_{ij}$ will be
related to the $(i, j)$ element of its inverse matrix.
\begin{eqnarray}
        \mathcal{C}_{ij}
                &\equiv&
        r_{\Pi_{\bot\{\Delta_{k\neq i, j}\}} \Delta_i \Pi_{\bot\{\Delta_{k\neq i, j}\}} \Delta_j }
                \equiv
        \frac{ (\Pi_{\bot\{\Delta_{k\neq i, j}\}} \VEC{\Delta}_i)^T 
		(\Pi_{\bot\{\Delta_{k\neq i, j}\}} \VEC{\Delta}_j ) }
        { \| \Pi_{\bot\{\Delta_{k\neq i, j}\}} \VEC{\Delta}_i \| \; \| \Pi_{\bot\{\Delta_{k\neq i, j}\}} \VEC{\Delta}_j \| }
                =
                - \; \frac{ (C^{-1})_{ij} }
                        {
                         \sqrt{ (C^{-1})_{ii} (C^{-1})_{jj} }
                        }
                \label{eq: partial_correlation_coefficient_matrix}
\end{eqnarray}

\subsubsection{Characteristic variables indicating coevolution between sites}

Characteristic changes accompanied by substitutions whose correlations indicate
coevolution between sites have been employed 
as $\Delta_{\kappa, \lambda}$ in \Eq{\ref{eq: mean_of_delta}}:
(1) occurrence of amino acid substitution 
(
$\Delta^{s}_{\kappa, \lambda} \equiv 1 - \delta_{a_\kappa, a_\lambda}$
),
(2) changes of side chain volume,	
(3) side chain charge, 	
(4) hydrogen-bonding capability accompanied by an amino acid substitution,
and (5) others\CITE{M:13}.
Then, a coevolution score $\rho_{ij}$ based on the partial correlation coefficient 
is defined for each characteristic. 
For example, in the case of concurrent substitutions between sites, 
$\rho^s_{ij} \equiv \max \; (\; \mathcal{C}^s_{ij}, \; 0 \; )$, because
the direct correlation of substitutions must be positive.
For volume, charge, and hydrogen-bonding capability changes, 
$\rho^x_{ij} \equiv \max ( - \text{sgn} \mathcal{C}^x_{ij}(| \rho^s_{ij} \mathcal{C}^x_{ij}|)^{1/2}, 0 )$
with $x =$ volume, charge, or hydrogen-bonding capability. 
Then, the total score was defined as $\rho_{ij} \equiv  \max (\rho^s_{ij}, \rho^v_{ij}, \rho^c_{ij}, \rho^h_{ij}, \ldots ) $;
refer to \CITE{M:13} for detail.
 
} 

\normalsize

\renewcommand{\url}[1]{}
\newcommand{\urlprefix}{}

\bibliographystyle{spmpsci}
\bibliography{jnames_with_dots,MolEvol,Protein,Bioinfo,SM}


\end{document}